\documentclass[twocolumn,twocolappendix]{aastex631}

\def\w0458{WISE-0458}
\def\Mjup{\ensuremath{M_\textrm{Jup}}}
\def\Rjup{\ensuremath{R_\textrm{Jup}}}

\usepackage[version=4]{mhchem}
\begin{document}

\title{\ce{HCN} and \ce{C2H2} in the atmosphere of a T8.5+T9 brown dwarf binary}

\author[0000-0003-0593-1560]{Elisabeth C. Matthews}
\affiliation{Max-Planck-Institut für Astronomie, Königstuhl 17, D-69117 Heidelberg, Germany}
\correspondingauthor{Elisabeth C. Matthews}
\email{matthews@mpia.de}

\author[0000-0003-4096-7067]{Paul Molli\`ere}
\affiliation{Max-Planck-Institut für Astronomie, Königstuhl 17, D-69117 Heidelberg, Germany}

\author[0009-0001-2738-2489]{Helena K\"uhnle}
\affiliation{Institute of Particle Physics and Astrophysics, 
ETH Zürich, Wolfgang-Pauli-Strasse 27, 8093, Zürich, Switzerland}

\author[0000-0001-8718-3732]{Polychronis Patapis}
\affiliation{Institute of Particle Physics and Astrophysics, 
ETH Zürich, Wolfgang-Pauli-Strasse 27, 8093, Zürich, Switzerland}

\author[0000-0001-8818-1544]{Niall Whiteford}
\affiliation{Department of Astrophysics, American Museum of Natural History, New York, NY 10024, USA}

\author[0000-0001-9992-4067]{Matthias Samland}
\affiliation{Max-Planck-Institut für Astronomie, Königstuhl 17, D-69117 Heidelberg, Germany}

\author{Pierre-Olivier Lagage}
\affiliation{Université Paris-Saclay, Université Paris Cité, CEA, CNRS, AIM, F-91191 Gif-sur-Yvette, France}

\author[0000-0002-5462-9387]{Rens Waters}
\affiliation{Department of Astrophysics/IMAPP, Radboud University, PO Box 9010, 6500 GL Nijmegen, The Netherlands}
\affiliation{SRON Netherlands Institute for Space Research, Niels Bohrweg 4, NL-2333 CA Leiden, the Netherlands}

\author[0000-0002-8163-4608]{Shang-Min Tsai}
\affiliation{Department of Earth and Planetary Sciences, University of California, Riverside, CA, USA}

\author{Kevin Zahnle}
\affiliation{Space Science Division, NASA Ames Research Center, Moffett Field, CA 94035, USA}
\affiliation{Virtual Planetary Laboratory, University of Washington, Seattle, WA 98195, USA}

\author[0000-0001-9818-0588]{Manuel Guedel}
\affiliation{Department of Astrophysics, University of Vienna, Türkenschanzstr. 17, 1180 Vienna, Austria}
\affiliation{Institute of Particle Physics and Astrophysics, 
ETH Zürich, Wolfgang-Pauli-Strasse 27, 8093, Zürich, Switzerland}

\author[0000-0002-1493-300X]{Thomas Henning}
\affiliation{Max-Planck-Institut für Astronomie, Königstuhl 17, D-69117 Heidelberg, Germany}

\author[0000-0002-1368-3109]{Bart Vandenbussche}
\affiliation{Institute of Astronomy, KU Leuven, Celestijnenlaan 200D, 3001 Leuven, Belgium}

\author[0000-0002-4006-6237]{Olivier Absil}
\affiliation{STAR Institute, Université de Liège, Allée du Six Août 19c, 4000 Liège, Belgium}

\author[0000-0003-2820-1077]{Ioannis Argyriou}
\affiliation{Institute of Astronomy, KU Leuven, Celestijnenlaan 200D, 3001 Leuven, Belgium}

\author[0000-0002-5971-9242]{David Barrado}
\affiliation{Centro de Astrobiología (CAB), CSIC-INTA, ESAC Campus, Camino Bajo del Castillo s/n, 28692 Villanueva de la
Cañada, Madrid, Spain}

\author[0000-0001-6492-7719]{Alain Coulais}
\affiliation{LERMA, Observatoire de Paris, Université PSL, CNRS, Sorbonne Université, Paris, France }
\affiliation{Université Paris-Saclay, Université Paris Cité, CEA, CNRS, AIM, F-91191 Gif-sur-Yvette, France}

\author[0000-0001-9250-1547]{Adrian M. Glauser}
\affiliation{Institute of Particle Physics and Astrophysics, 
ETH Zürich, Wolfgang-Pauli-Strasse 27, 8093, Zürich, Switzerland}

\author[0000-0003-3747-7120]{Goran Olofsson}
\affiliation{Department of Astronomy, Stockholm University, AlbaNova University Center, 10691 Stockholm, Sweden}

\author[0000-0002-0932-4330]{John P. Pye}
\affiliation{School of Physics \& Astronomy, Space Park Leicester, University of Leicester, 92 Corporation Road, Leicester, LE4 5SP, UK}

\author[0000-0002-2352-1736]{Daniel Rouan}
\affiliation{LESIA, Observatoire de Paris, Université PSL, CNRS, Sorbonne Université, Université de Paris Cité, 5 place Jules Janssen, 92195 Meudon, France}

\author[0000-0001-9341-2546]{Pierre Royer}
\affiliation{Institute of Astronomy, KU Leuven, Celestijnenlaan 200D, 3001 Leuven, Belgium}

\author[0000-0001-7591-1907]{Ewine F. van Dishoeck}
\affiliation{Leiden Observatory, Leiden University, PO Box 9513, 2300 RA Leiden, The Netherlands}

\author[0000-0002-2110-1068]{T.P. Ray}
\affiliation{School of Cosmic Physics, Dublin Institute for Advanced Studies, 31 Fitzwilliam Place, Dublin, D02 XF86, Ireland}

\author[0000-0002-3005-1349]{Göran Östlin}
\affiliation{Department of Astronomy, Oskar Klein Centre, Stockholm University, 106 91 Stockholm, Sweden}

\begin{abstract}

T-type brown dwarfs present an opportunity to explore  atmospheres teeming with molecules such as \ce{H2O}, \ce{CH4} and \ce{NH3}, which exhibit a wealth of absorption features in the mid-infrared. With JWST, we can finally explore this chemistry in detail, including for the coldest brown dwarfs that were not yet discovered in the Spitzer era. This allows precise derivations of the molecular abundances, which in turn informs our understanding of vertical transport in these atmospheres and can provide clues about the formation of cold brown dwarfs and exoplanets. This study presents the first JWST/MRS mid-IR spectrum (R$\sim$1500-3000) of a T-dwarf: the T8.5+T9 brown dwarf binary WISE J045853.90+643451.9. We fit the spectrum using a parameterized $P$-$T$ profile and free molecular abundances (i.e., a retrieval analysis), treating the binary as unresolved. We find a good fit with a cloud-free atmosphere and identify \ce{H2O}, \ce{CH4} and \ce{NH3} features. Moreover, we make the first detections of \ce{HCN} and \ce{C2H2} (at 13.4$\sigma$ and 9.5$\sigma$ respectively) in any brown dwarf atmosphere. The detection of \ce{HCN} suggests intense vertical mixing ($K_{zz}\sim10^{11}$\,cm$^2$s$^{-1}$), challenging previous literature derivations of $K_{zz}$ values for T-type brown dwarfs. Even more surprising is the \ce{C2H2} detection, which cannot be explained with existing atmospheric models for isolated objects. This result challenges model assumptions about vertical mixing, and/or our understanding of the \ce{C2H2} chemical network, or might hint towards a more complex atmospheric processes such as magnetic fields driving aurorae, or lightning driving ionization. These findings open a new frontier in studying carbon chemistry within brown dwarf atmospheres.

\end{abstract}


\section{Introduction}

Brown dwarfs provide a laboratory for studying the physics and chemistry of cold atmospheres. Their atmospheres are similar to those of low-irradiation exoplanets, though with some differences due to their higher mass and surface gravity. Further, their atmospheres are relatively accessible with spectroscopic characterization, without some of the observational complications that arise for exoplanets (which are typically overwhelmed by the light of a much brighter host-star). Brown dwarfs and exoplanets are both spectrally classified as late-M, L, T, and Y-type objects \citep{Cushing2006,Cushing2011}. As they cool, more molecules are able to form in their atmospheres, leading to increasingly complex atmospheric chemistry and corresponding emission spectra, though at the very coldest temperatures chemistry is kinetically inhibited.

T-type objects have strong water and ammonia absorption features in the mid-IR \citep{Suarez2022}. In cool T-dwarfs with high-surface gravity, carbon is predicted to be primarily in the form of \ce{CH4}, with the \ce{CO} atmospheric content correlating with the vertical mixing strength \citep[e.g.][]{Zahnle2014}. Mid- to late-T type brown dwarfs should have relatively clear atmospheres, since they are cold enough that silicate clouds in the atmosphere have sunk deep into the atmosphere, but warm enough that water- and ammonia-ice clouds are not expected.

\citet{Zahnle2014} predicted that \ce{HCN} might be present for high-gravity brown dwarfs with very strong vertical mixing (parametrized through an eddy diffusion coefficient ($K_{zz}$, with very strong vertical mixing corresponding to $K_{zz}\gtrapprox10^{10}$\,cm$^2$s$^{-1}$). \ce{HCN} is only produced through chemical reactions at higher temperatures and pressures, but can become ``quenched" (i.e., the mixing timescale becomes faster than the chemical reaction timescale, leading to a vertically constant abundance, \citealt{Prinn1977}) and reach observable abundances higher in the atmosphere. However, \citet{Mukherjee2024} demonstrated that vertical mixing is relatively weak in field brown dwarfs. They studied the spectra of a sample of nine early- to late-T dwarfs observed with \textit{Spitzer} and \textit{AKARI}, and derived typical vertical mixing values in the atmosphere of $K_{zz}\sim10^{1}-10^{4}$\,cm$^2$s$^{-1}$. A more recent study of a Y0-dwarf with JWST NIRSpec/PRISM and MIRI/LRS found a preference for models with $\log(K_{zz}) \geq 4$\,cm$^2$s$^{-1}$ \citep{Beiler2023} -- but that work highlights that the strong vertical mixing is required to match the deep CO and NH$_3$ features. This strong mixing is inconsistent with the PH$_3$ non-detection in that object, possibly indicating the need for a more complex vertical mixing prescription (though see \citealt{Beiler2024b} for additional discussion of the PH$_3$ non-detection).

\ce{C2H2} was not predicted in the atmospheres of non-irradiated brown dwarfs, but both \ce{HCN} and \ce{C2H2} should be formed by photo-dissociation in irradiated objects \citep[e.g.][]{Sharp2004,Moses2011}, and in these cases they are a tracer of a carbon-rich atmosphere \citep{Venot2015}. In the case that \ce{C2H2} is produced by photo-dissociation, it should appear primarily at very high altitudes/low pressures (e.g. \citealt{Venot2015} predicted \ce{C2H2} abundance to exceed a molar fraction of $10^{-12}$ at above $\sim10^{-3}$\,bar, and peak at a molar fraction of $\sim10^{-5} - 10^{-4}$ at $\sim10^{-6}$\,bar in a 500\,K atmosphere with external radiation.

JWST provides the opportunity to characterize these T-dwarfs to a level of detail not previously possible. The medium resolution spectrograph \citep[MRS][]{Wells2015,Argyriou2023} of the JWST/MIRI instrument \citep{Rieke2015,Wright2015}  provides high-sensitivity spectra at R$\sim$1500--3500 from 4.9\,\micron~to 27.9\,\micron. For cold atmospheres, this unveils a rich forest of molecular absorption features. Throughout this work, we will refer to the 1-5\,\micron~wavelength range as the near-IR and to $\lambda>5$\,\micron~as the mid-IR.

No MIRI/MRS spectra of T-type brown dwarfs have to date been published, though a few publications have studied L-type \citep{Miles2023} and Y-type \citep{Barrado2023,Kuehnle2024} objects. In this work we present the first MIRI/MRS T-dwarf spectrum: observations of the brown dwarf binary WISE J045853.90+643451.9 (hereafter \w0458), providing an opportunity to search for new chemistry and to constrain the atmospheric structure of a cold object. \w0458 was the first ultra-cool brown dwarf identified by the WISE mission \citep{Mainzer2011,Kirkpatrick2011}. It was originally classified as T9 or T8.5 by \citet{Mainzer2011} and \citet{Cushing2011} respectively, and has strong near-IR \ce{H2O} and \ce{CH4} absorption bands, as well as tentative evidence for \ce{NH3} absorption in the near-IR \citep{Burgasser2012}. The system is at a distance of $9.24^{+0.14}_{-0.15}$~pc \citep{Leggett2019}, and has no identified markers of relative youth, suggesting it is a field-age object with high surface gravity. 

\w0458 was soon resolved as a binary brown dwarf in Keck high-resolution images by \citet{Gelino2011}, who found spectral types T8.5 and T9 and temperatures $\sim$600\,K and $\sim$500\,K respectively for the two components. The pair has a semi-major axis of $5.0^{+0.3}_{-0.6}$\,au  ($540^{+40}_{-70}$\,mas), and at the epoch of JWST observation were at a predicted mutual separation of 330$\pm$40\,mas, based on the orbit fit of \citet{Leggett2019}. This corresponds to a separation of 1.3$\times$ the diffraction limit at 5\,\micron, and 0.45$\times$ the diffraction limit at 18\,\micron~\citep{Law2023}, i.e., the pair is barely resolved at the shortest MRS wavelengths, and unresolved at the longest. We therefore treat the pair as an unresolved binary throughout this work, and assume identical atmospheres for both objects; the similarity in spectral type and temperature between both objects justifies this approach. 

For any binary, in the extreme cases either (1) the flux of one object dominates, and models would constrain only the physical parameters of that object or (2) the objects are identical, in which case models can also constrain all the physical parameters of both objects, but will return a radius inflated by $\sqrt{2}$ (corresponding to twice the emitting area of one object). In the intermediate case (both objects contribute to the flux, but have different atmospheres) retrievals may be more degenerate and measured parameters can be biased. For \w0458, the pair of brown dwarfs have similar fluxes and temperatures: \citet{Leggett2019} find the pair to be $\sim$1mag different in brightness in the near-IR, and derive temperatures $\sim$600\,K and $\sim$500\,K respectively for the two atmospheres. If we infer that both formed via a star-formation pathway and at the same time and location, they should also have similar atmospheric compositions. For this work, we therefore approximate the spectrum as a single atmosphere and assume a common effective temperature and composition, though we intend to revisit this assumption in a future work. As the library of late-T and Y dwarfs with JWST spectra grows, it will be important to understand the biases that are introduced when modelling the atmospheres of a binary brown dwarf in this way -- this is likely particularly important in temperature ranges where the atmosphere is changing rapidly (e.g. the L-T transition and the T-Y transition).

The paper is structured as follows: we first present the observations and data reduction in Sections \ref{sec:obs} and \ref{sec:datareduction} respectively. In Section \ref{sec:spectrum} we qualitatively describe the mid-IR spectrum of \w0458, and in Section \ref{sec:retrievals} describe our retrieval analysis to constrain its atmospheric structure and composition. In section \ref{sec:newmolecules} we discuss the molecular content of the atmosphere, and in particular the detections of \ce{HCN} and surprisingly of \ce{C2H2}, and the implications of these detections. Finally, we present some more general discussion and concluding remarks in Sections \ref{sec:discussion} and \ref{sec:conclusions} respectively.

\section{Observations}
\label{sec:obs}

We observed \w0458~with JWST/MIRI instrument on 21~Nov~2022, using the MRS. Observations were collected as part of the European MIRI GTO consortium. NIRSpec observations of \w0458~were also collected by another GTO team, and will be presented in a future paper (Lew et al.~in prep). To reduce overheads, both the MRS and NIRSpec observations were collected through the same program (PID 1189). 

The MRS has four channels (1 through 4) that cover consecutive wavelength regions and are observed simultaneously; together these four channels span 4.9\,\micron~to 27.9\,\micron. These channels can be observed with any one of three grating settings (referred to as short, medium and long or as A, B and C), each covering one-third of each channel, for a total of 12 subchannels with unique grating/channel combinations. For \w0458, we integrated for 2997s with each grating setting. At the longest wavelengths the brown dwarf signal becomes overwhelmed by thermal and background noise, and in this work we only consider channels 1--3 (4.9\,\micron~to 18.0\,\micron). Observations were performed in the FASTR1 readout mode and in a 2-point point-source optimized dither pattern to provide better sampling of the PSF; this dither also has the advantage of providing a contemporaneous observation that can be used for background subtraction. For each of the three grating settings, we collected one integration per dither position with 180 groups per integration.

\section{Data Reduction}
\label{sec:datareduction}

Data reduction was performed using the public \texttt{jwst} pipeline\footnote{\url{https://jwst-pipeline.readthedocs.io/en/latest/}} (version 1.12.5; \citealt{Bushouse_jwst}) with the corresponding CRDS file version 11.17.10 and context files jwst$\_$1183.pmap. 
The reduction consists of three distinct stages. The first stage of the pipeline calculates the rate of photoelectric charge accumulation from the ramp files. The second stage, after assigning a world coordinate system to the data, includes applying several calibrations such as correcting for the flat field, the characterized scattered light internal to the MIRI detectors, the fringes in the spectral direction, and the photometric flux calibration \citep[for additional discussion of the flux calibration of MIRI/MRS, see][]{Law2025}. After this step, the background is removed by subtracting the two dithers from each other (i.e.~a nod subtraction). 
The final stage uses the generated and corrected detector files to produce image cubes. The algorithm to project from the 2D to 3D cube in this case is the ‘drizzle’ weighting algorithm \citep{Law2023}. Before generating the cubes, an outlier detection is performed to mask out cosmic rays and similar residual outliers. 

From the cube, the one-dimensional spectrum is extracted by placing an aperture around the source. For \w0458, we center the aperture around the brightest point in the collapsed datacube for each channel, using the built-in \texttt{ifu$\_$autocen()} function. We use a radius of twice the analytical FWHM around the source, which varies between $\sim$0.3'' and $\sim$1'' over the MIRI wavelength range; this ensures flux from both brown dwarfs (separation $\sim$330\,mas) is captured within the aperture. The spectrum is extracted with the \texttt{extract$\_$1d()} function built into the \texttt{jwst} pipeline. 

\subsection{Residual background effects}

With the pipeline data reduction as described above, the backgrounds are imperfectly corrected, and large-scale systematics remain in the 3D cubes (see Appendix \ref{app:backgrounds} for a demonstration, and \citealt{Argyriou2023} for detailed discussion of the MIRI in-flight performance and systematics).
We therefore developed a custom, data-driven post-processing treatment to correct background systematics in the 3D cubes, which is described in detail in Appendix \ref{app:backgrounds}. 

\section{The mid-IR spectrum of \w0458}
\label{sec:spectrum}

\begin{figure*}[ht]
    \centering
    \includegraphics[width=.94\textwidth]{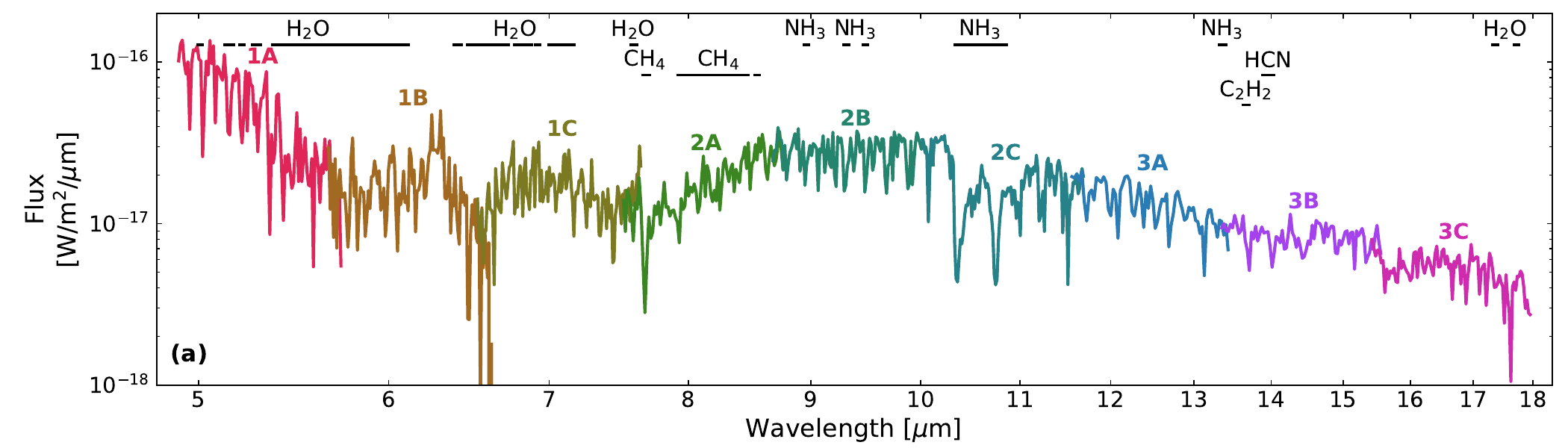}
    \includegraphics[width=.94\textwidth]{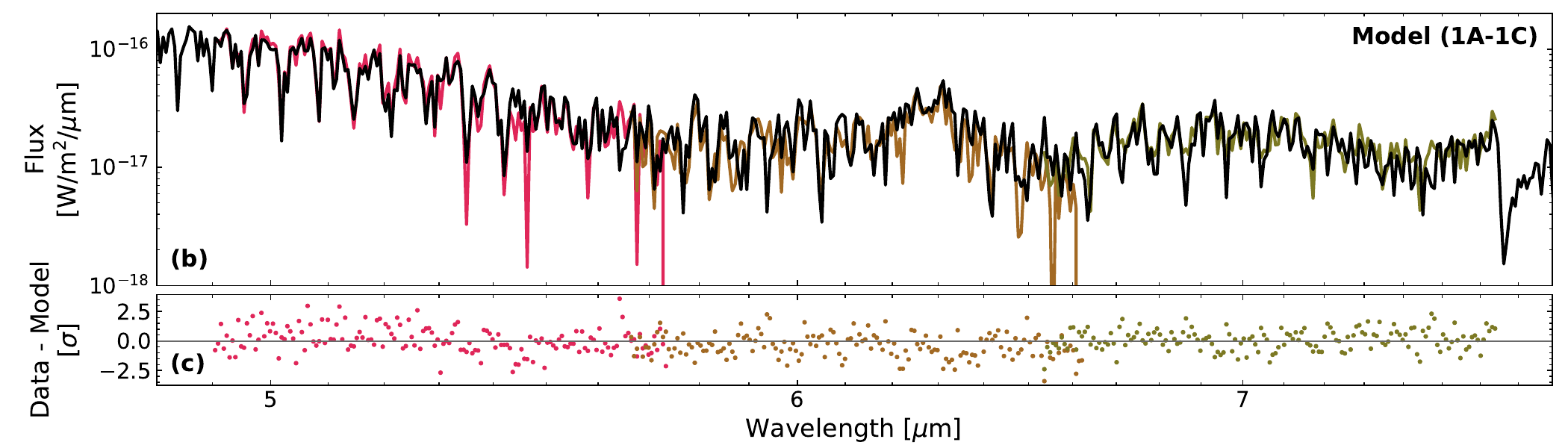}
    \includegraphics[width=.94\textwidth]{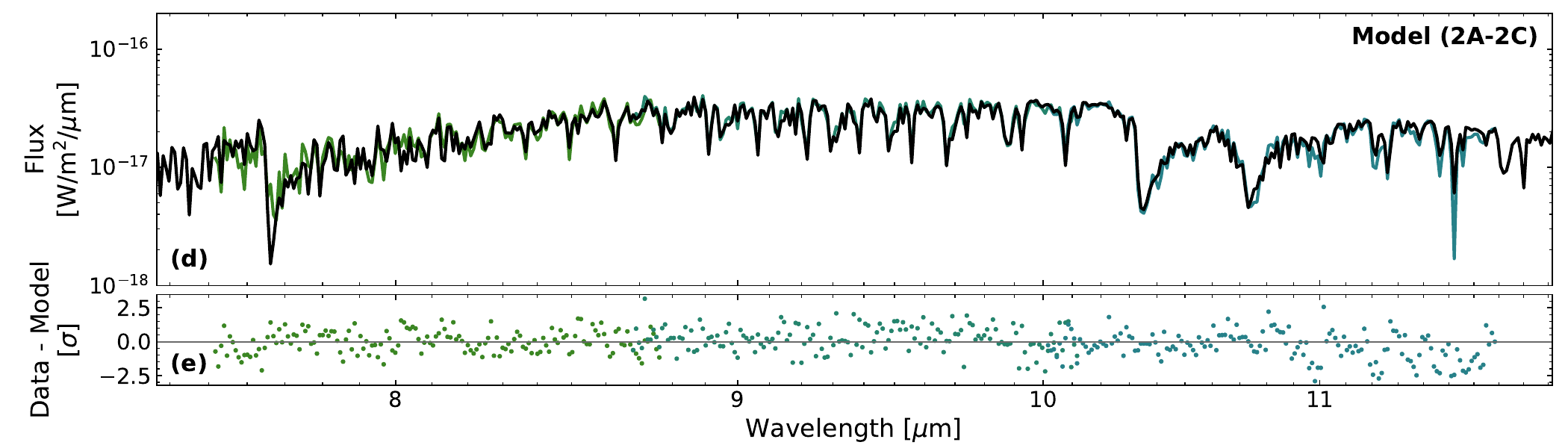}
    \includegraphics[width=.94\textwidth]{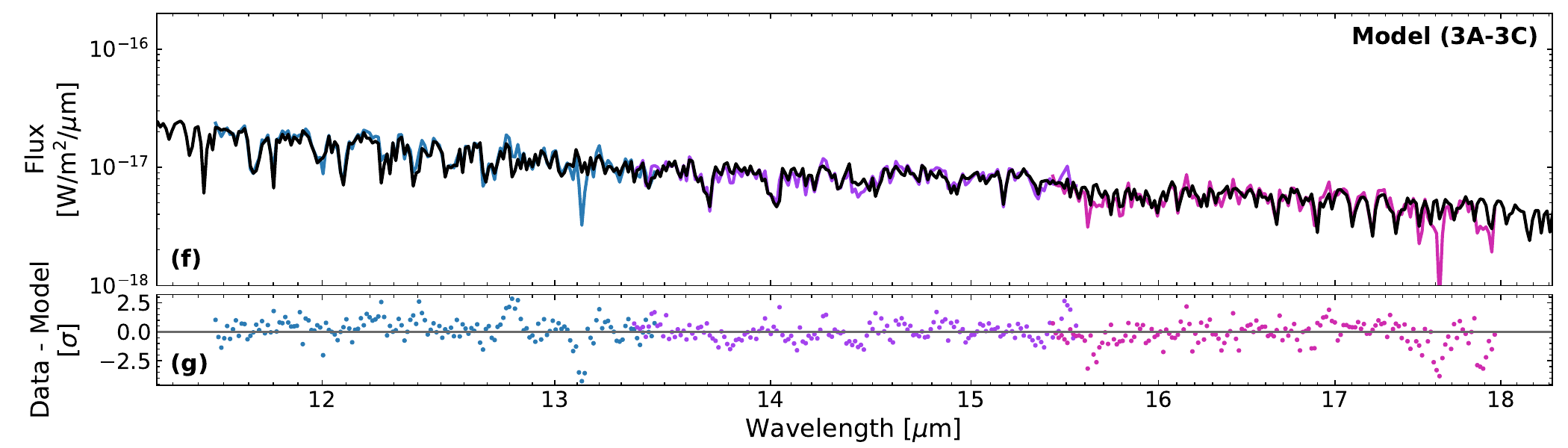}
    \caption{The MIRI/MRS spectrum of \w0458, and our best-fit model. Panel (a) shows the full spectrum, binned to 1/10th the original spectral resolution for visual clarity. Colors represent each individual subchannel of the MRS detector (labelled 1A-3C), and we also label the locations of prominent molecular absorption features in this brown dwarf atmosphere. Panels (b), (d), (f) show the best-fit model (black) compared to the data for each channel (at R=1,000). Our model is a free retrieval using \texttt{pRT} (see Section \ref{sec:retrievals}), including a parameterized pressure-temperature profile and 7 molecular species; 5 molecules are confidently detected (\ce{H2O}, \ce{CH4}, \ce{NH3}, \ce{HCN} and \ce{C2H2}). Panels (c), (e) and (g) show the residuals, taking the 10$^b$ factor into account.}
    \label{fig:mainspectrum_and_model}
\end{figure*}

The full MIRI/MRS spectrum of \w0458 is shown in Figure~\ref{fig:mainspectrum_and_model}, with the key molecular absorptions highlighted. \ce{H2O}, \ce{CH4} and \ce{NH3} all have a forest of molecular lines in the mid-IR. In particular, \ce{H2O} dominates the shape of the spectrum between 5--7\,\micron~and a very clear \ce{NH3} feature is visible between 10--11\,\micron. \ce{CH4} and \ce{NH3} also contribute to the forest of molecular lines beyond $\sim$7\,\micron.

\begin{figure*}[ht]
    \centering
    \includegraphics[width=.99\textwidth]{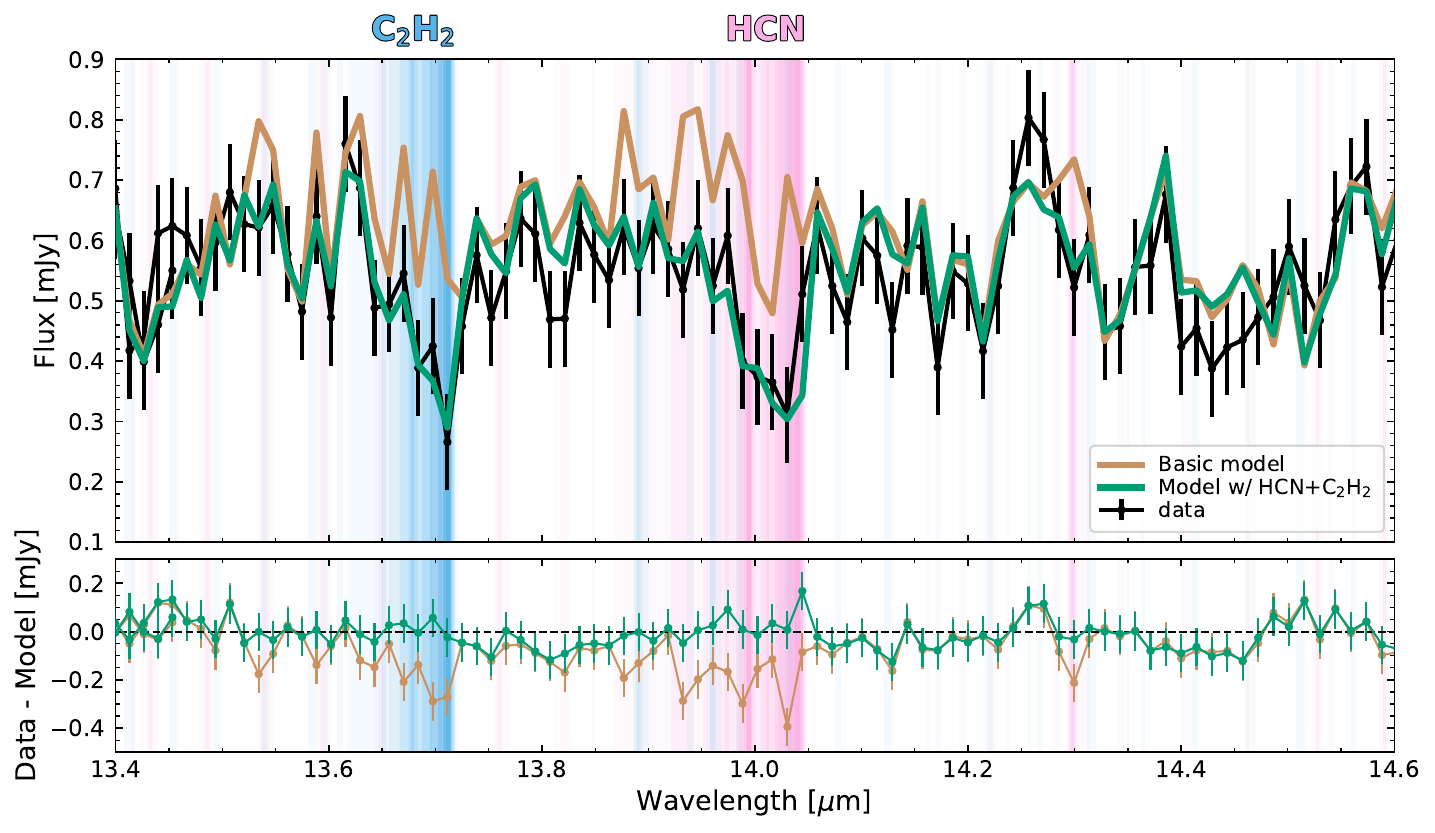}
    \caption{Comparison of the data and best-fit models with and without including the molecules \ce{HCN} and \ce{C2H2}. Here we show our retrieved spectra for a ``basic" model with only \ce{H2O}, \ce{CH4}, \ce{NH3}, CO and CO$_2$ (brown), and a model that also includes \ce{HCN} and \ce{C2H2} (green), alongside the data (black, including our retrieved uncertainty inflation). The lower panel shows the residuals (data - model) for the both models, and shading in the background of both panels indicates regions of high opacity for \ce{HCN} (pink) and \ce{C2H2} (blue). The plot shows only the 13.4-14.6\,\micron~region of the spectrum where \ce{HCN} and \ce{C2H2} absorption features are significant (see also Fig.~\ref{fig:hcnopacity}), but these species have a small continuum effect across more of the mid-IR.}
    \label{fig:hcn_c2h2}
\end{figure*}

Absorption features from \ce{HCN} and \ce{C2H2} are visible at 13.75\,\micron~and 14\,\micron~respectively (Figure~\ref{fig:hcn_c2h2}; these detections are both confirmed with a retrieval analysis (Section \ref{sec:hcn}; \ref{sec:c2h2}). Neither species has previously been observed in a brown dwarf atmosphere. While there are several mid-IR brown dwarf spectra measured by Spitzer \citep{Suarez2022}, we here present the \textit{first} T-dwarf spectrum with JWST/MIRI/MRS, and therefore the first T-dwarf spectrum with this resolution and SNR in the mid-IR, meaning that we are more sensitive than any previous observations to \ce{HCN} and \ce{C2H2}. Among the heavily-irradiated exoplanets, detections of both \ce{HCN} and \ce{C2H2} have been claimed in the atmosphere of HD209458b \citep{Hawker2018,Giacobbe2021}, though the detections of both species are disputed based on JWST data \citep{Xue2024} and on high-resolution data at $\sim$3\,\micron~\citep{Blain2024}. Finally, we note that both species are commonly seen as emission features in disks of young stars and brown dwarfs, with Spitzer \citep{Pascucci2013} and more recently JWST \citep{Henning2024,Tabone2023,Arabhavi2024}.

\begin{figure*}[ht]
    \centering
    \includegraphics[width=1.\textwidth]{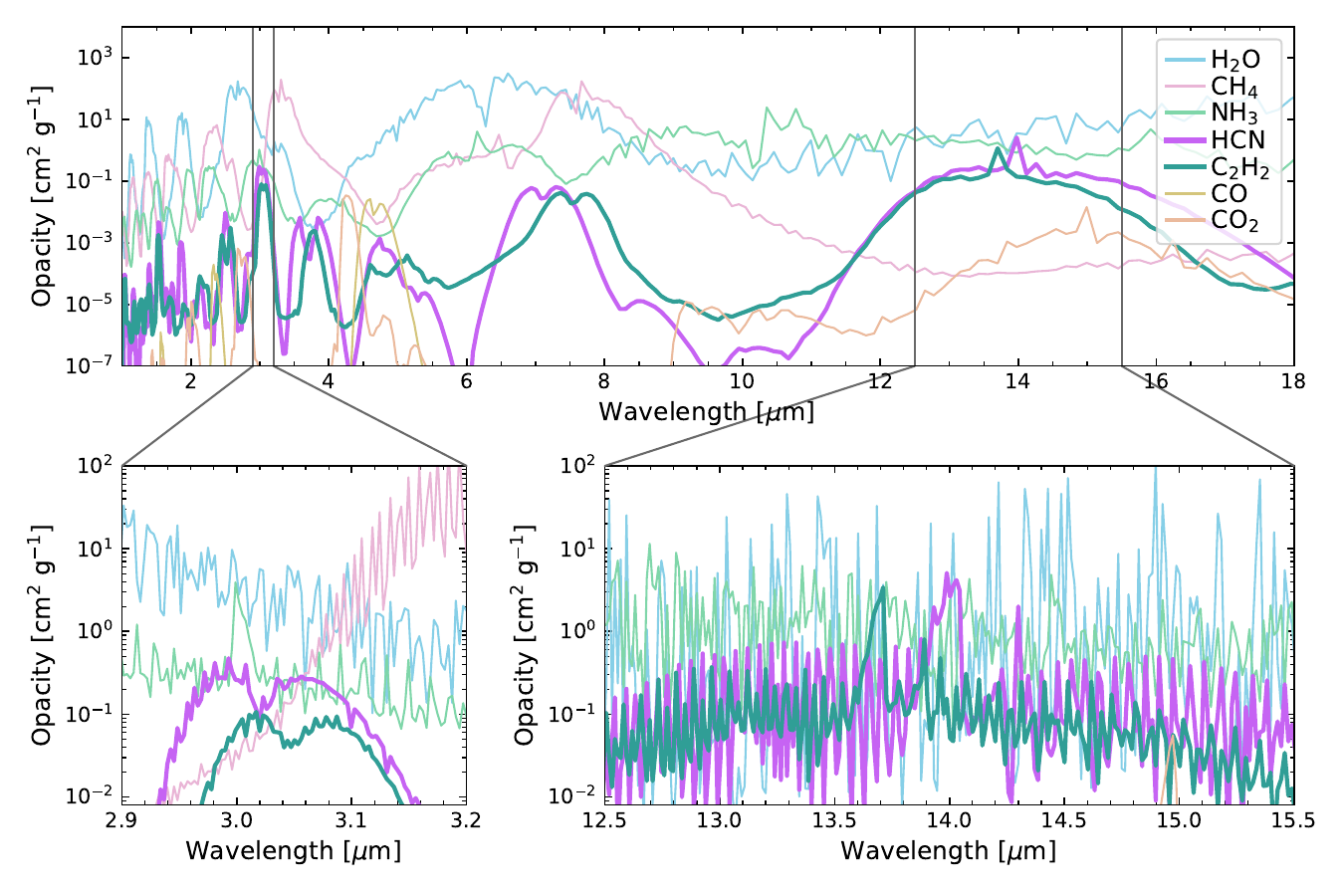}
    \caption{Opacities of molecules in our best-fit model, with abundances matching the best-fit value (for \ce{H2O}, \ce{CH4}, \ce{NH3}, \ce{HCN} and \ce{C2H2}) and 2$\sigma$ upper limits (for CO and CO$_2$) from our best-fit model. Opacities in the upper panel are at R=100 and in the lower panel are at R=1000 (matching the resolution used in our \texttt{pRT} retrievals). In the lower panel we highlight the regions where \ce{HCN} and \ce{C2H2} absorb most strongly in the near- and mid-IR. For the \w0458 abundances, both species are only detectable in the mid-IR around 14\,\micron~in the \w0458 atmosphere. With sufficiently high abundances (higher than those observed here), both species might become detectable in the near-IR, at 3\,\micron~and 1.5\,\micron.}
    \label{fig:hcnopacity}
\end{figure*}

\section{Retrieval Modelling Approach}
\label{sec:retrievals}

We modeled the spectrum of \w0458~using a retrieval analysis. For simplicity, we assumed both brown dwarfs to have the same $P$-$T$ structure and atmospheric composition; this is a valid approximation since the spectral types of each component are very similar (T8.5+T9, \citealt{Leggett2019}). We used the \texttt{petitRADTRANS}\footnote{\url{https://petitradtrans.readthedocs.io}} retrieval package (hereafter \texttt{pRT}; \citealt{Molliere2019,Nasedkin2024}), which allows for emission spectra to be generated, and used a nested sampling technique \citep{Skilling2004} as implemented in \texttt{pyMultiNest} \citep{Feroz2008,Buchner2014} to probe the parameter space and derive posterior distributions for atmospheric properties.

We binned the MRS data from a native resolution of between 1,500 and 3,500 to $\lambda$/$\Delta\lambda$ = 1,000 when running retrievals. This allows for an efficient opacity treatment with correlated-$k$ opacities (rather than line-by-line opacities), where 1000 is \texttt{pRT}’s finest wavelength spacing in correlated-$k$, and thereby minimizes computation time. We modeled the atmosphere with a free pressure-temperature ($P$-$T$) profile with 10 temperature nodes, spaced equidistantly between $10^{-6}$ and $10^{3}$ bar in log-space. The temperature at each `node' is a free-parameter, and the full $P$-$T$ profile is calculated by quadratic interpolation between these nodes. This led to a ``wiggly'' $P$-$T$ profile, with several kinks in the posterior distribution, so we therefore also regularized the profile following \citet{Line2015} \citep[see also][]{Barrado2023}. Briefly, this approach penalizes rough profiles (defined based on their $\textrm{d}^2\log{T}/\textrm{d}\log{P}^2$), and includes a free parameter $\gamma$ that dictates the degree of smoothing. We also inflated the pipeline-derived observational uncertainties, and retrieved an inflation term following \citet{Line2015}, namely, we calculate inflated uncertainties for each datapoint following the formula $\sigma = \sqrt{\sigma_\textrm{red}^2 + 10^b}$, where $b$ is a free parameter in the retrieval and $\sigma_\textrm{red}$ is the observational uncertainty on each datapoint, derived from the \texttt{jwst} pipeline and binned to R=1000 assuming uncorrelated errors. This allows a converged fit even in the presence of underestimated uncertainties in the data and/or missing physics in the model. A more physically accurate model is then expected to have a lower $b$ parameter and smaller inflated uncertainties; this approach also accounts for the underestimated uncertainties in the \texttt{jwst} pipeline (see below).

Our basic model for the retrieval includes \ce{CH4}, CO$_2$, CO, \ce{H2O} and \ce{NH3}, the main species that are expected in cold atmospheres.  We assumed all molecules to have vertically constant abundances that were retrieved freely, which is a common assumption for brown dwarf modelling, and is likely accurate across the pressures we probe with MRS spectra. We did not include clouds in our model. Line-lists for these species are from HITEMP for \ce{CO}, \ce{CO2} \citep{Rothman2010} and \ce{CH4} \citep{Hargreaves2020}, and from ExoMOL for \ce{H2O} \citep{Polyansky2018} and \ce{NH3} \citep{Coles2019}. Alongside these five abundance parameters, our model has 11 temperature/pressure free parameters (10 nodes and the $\gamma$ parameter used in the $P$-$T$ profile regularization), two additional physical parameters ($\log{g}$ and radius of the brown dwarf), and the $b$ parameter to inflate uncertainties, for a total of 19 free parameters in the basic model. We also tested for the presence of other molecular species, in each case adding the molecule individually to the retrieval and repeating the analysis. Line-lists for additional species are from HITRAN2012 for \ce{H2S} and \ce{C2H2} \citep{Rothman2013}; HITRAN2020 for \ce{^{15}NH3} and \ce{C2H6} \citep{Gordon2022}; ExoMol for \ce{C2H4} \citep{Mant2018} and \ce{PH3} \citep{Sousa-Silva2015}; and from \citet{Harris2006} for \ce{HCN}.

\section{Molecules in the \w0458 atmosphere}
\label{sec:newmolecules}

We found a well-fitting model that includes \ce{H2O}, \ce{CH4}, \ce{NH3}, \ce{HCN} and \ce{C2H2} with our retrieval approach. Neither \ce{CO} nor \ce{CO2} are detected; the non-detection of \ce{CO2} is consistent with models while the non-detection of \ce{CO} is expected only for relatively weak vertical mixing. This model is shown in Figure \ref{fig:mainspectrum_and_model}; no strong features remain unexplained with this model and there are no obvious wavelength-dependent systematics. There is some hint of \ce{CO2} in the models without \ce{HCN}, but this disappears when \ce{HCN} is included and we infer it is a false-positive detection. Aside from the sharp 14\,\micron~detection feature, \ce{HCN} has some non-negligible opacity between $\sim$12-16\,\micron~(Fig.~\ref{fig:hcnopacity}), and correspondingly slightly changes the broadband spectral shape throughout this region; in a model without \ce{HCN}, the retrieved \ce{CO2} abundance is elevated to explain this broadband spectral shape. We also see some correlation between the retrieved \ce{HCN} and \ce{C2H2} for the same reason: both have a similar impact on the broadband spectral shape between $\sim$12-16\,\micron, and in a model with only one of these species included, the abundance of that molecule is overestimated to better fit the broadband shape of the other molecule. A full table of posteriors, and a corner plot of molecular abundances, are included in Appendix \ref{app:posteriors_table_corner}.

The derived atmospheric C/O ratio for this retrieval is $0.35\pm0.03$ (driven mainly by the \ce{H2O} and \ce{CH4} abundances), and the atmospheric metallicity is $1.35^{+0.19}_{-0.15} \times$ solar, i.e., a slightly super-solar metallicity and slightly sub-solar C/O ratio. The derived effective temperature is $566^{+7}_{-6}$\,K. For the retrieval without \ce{HCN} and \ce{C2H2}, we find consistent C/O values ($0.33\pm0.03$) but a slightly higher metallicity and effective temperature ($1.71^{+0.31}_{-0.25}\times$ solar; $580\pm5$\,K). While these calculations formally include the CO and \ce{CO2} abundances returned by the retrieval, the abundance upper limits on both species are sufficiently constraining that they do not impact the derived C/O and metallicity.

In the following text we discuss in detail the newly detected molecules \ce{HCN} and \ce{C2H2}, and then briefly also mention several molecules that are not detected in the current spectrum.

\subsection{Hydrogen Cyanide (\ce{HCN}) detection}
\label{sec:hcn}

A clear \ce{HCN} absorption feature is visible at 14\,\micron~(Figure \ref{fig:hcn_c2h2}), as well as a smaller feature at 14.3\,\micron; in both cases the data clearly matches the absorption in the best-fit model from the retrieval. We find a Bayes factor of 37.0 between the model with and without \ce{HCN} (including \ce{C2H2} in both cases), corresponding to an \ce{HCN} detection significance of 13.4$\sigma$. 

In a theoretical study, \citet{Zahnle2014} predicted that \ce{HCN} would only be present in the atmospheres of brown dwarfs with very strong vertical mixing and a high surface gravity (corresponding to a compressed scale height), using a one-dimensional chemical kinetics code coupled with a physically-motivated $P$-$T$ profile. Our observed \ce{HCN} abundances are broadly consistent with the predicted \ce{HCN} abundances for their model with the strongest vertical mixing.
For a 600\,K model atmosphere with $\log{g}=5.0$ and metallicity$=1\times$ solar, \citet{Zahnle2014} predict a quench-point \ce{HCN} abundance (volume mixing ratio, hereafter VMR) of $7.8\times10^{-7}$ for their most vigorously mixed model with $K_{zz}=10^{11}$\,cm$^2$s$^{-1}$; in our retrievals we observe an VMR of $12^{+4}_{-3} \times 10^{-7}$. For a more modest value of $K_{zz}=10^{4}$\,cm$^2$s$^{-1}$, \citet{Zahnle2014} predict an \ce{HCN} VMR of $1.1\times10^{-9}$, well below the abundance observed here. These models have predicted \ce{NH3} VMRs of $2.2\times10^{-5}$ and $2.0\times10^{-5}$; both values are $\sim2\sigma$ below our retrieved the \ce{NH3} abundance ($3.0^{+0.5}_{-0.4} \times 10^{-5}$). The \ce{HCN}/\ce{NH3} abundance ratio is less sensitive to the surface gravity than the \ce{HCN} abudance, and as such could provide a better proxy for the vertical mixing in the atmosphere. For \w0458, both \ce{HCN} and \ce{NH3} are slightly enhanced relative to the \citet{Zahnle2014} predictions, consistent with the slightly higher surface gravity of \w0458 relative to their models; the \ce{HCN}/\ce{NH3} abundance ratio is as expected for a high-gravity object with strong vertical mixing ($K_{zz}=10^{11}$\,cm$^2$s$^{-1}$). However, these results indicate much stronger vertical mixing than previously inferred for T- and Y-dwarfs \citet{Mukherjee2024,Beiler2023,Xuan2024}.

A strongly enhanced metallicity could also promote a higher \ce{HCN} abundance in the brown dwarf atmospheres. \citet{Zahnle2014} include model atmospheres with $3\times$ solar metallicity, but do not simultaneously model a temperature of 600K with an atmospheric metallicity of $3\times$ solar. At 500K, their model with $3\times$ solar metallicity has $\sim5\times$ the \ce{HCN} content as the model with $1\times$ solar metallicity (values are $1.2\times10^{-6}$ and $2.0\times10^{-7}$ and respectively). This suggests that even a significant enhancement in metallicity would be insufficient to reconcile our observed \ce{HCN} abundance with that predicted for a $K_{zz}=10^{4}$\,cm$^2$s$^{-1}$ atmosphere. Further, our derived atmospheric metallicity ($1.35^{+0.19}_{-0.15}\times$ solar) is only minimally enhanced relative to solar, and cannot solve the conflict between our high-\ce{HCN} abundance and the previously claimed weak vertical mixing. Similarly, high C/O could promote production of \ce{HCN}, but the observed C/O ratio of \w0458 is not in this regime.

\ce{CH4} is favored over CO in cool T-dwarfs with high surface gravity, regardless of the vertical mixing, though the highest vertical mixing prescriptions CO abundance is increased almost to the level of \ce{CH4} \citep{Zahnle2014}. In \w0458, we detect \ce{CH4} and not \ce{CO}, which is more in-line with a lower $K_{zz}$ atmosphere than the \ce{HCN} abundance suggests. Generating the high \ce{HCN} abundances we observe without also generating detectable levels of \ce{CO} is challenging for models. Upcoming analysis of NIRSpec data for this target (Lew et al.~in prep) will significantly improve the CO constraints for \w0458, which primarily absorbs shortwards of 5\,\micron~(Fig.~\ref{fig:hcnopacity}), and may provide further clues about the atmospheric mixing in \w0458.

\subsection{Acetylene (\ce{C2H2}) detection}
\label{sec:c2h2}

\ce{C2H2} is readily visible in the \w0458 spectrum, most notably with a clear absorption feature at 13.7\,\micron~(Figure \ref{fig:hcn_c2h2}). In our retrieval analysis for \w0458, the best-fit model with \ce{C2H2} is favored with a Bayes factor of 18.4 over the model without (including \ce{HCN} in both cases), corresponding to a \ce{C2H2} detection significance of 9.5$\sigma$. As with \ce{HCN}, the only clear absorption features appear around 13-14\,\micron, but there is also some impact on the continuum flux level at other wavelengths. \ce{C2H2} also has a prominent 7.7\,\micron~feature, as seen in some protoplanetary disk spectra \citep[e.g][]{Tabone2023}, but in the \w0458 spectrum this feature is overwhelmed by the strong \ce{H2O}, \ce{CH4} and \ce{NH3} absorption.

\ce{C2H2} has not previously been seen in a brown dwarf atmosphere, and is an unexpected discovery. At first glance, the presence of \ce{C2H2} is consistent with the detection of \ce{HCN}: both species are indicators of disequilibrium chemistry, and likely form together deeper in the brown dwarf atmosphere. However, several brown dwarf model atmospheres \citep{Marley2021,Mukherjee2024} include \ce{C2H2}, and suggest that the atmospheric abundance of \ce{C2H2} should be extremely low. Our observed abundance ($2.7^{+1.4}_{-1.1}\times10^{-7}$) of \ce{C2H2} in the \w0458 atmosphere cannot be explained within a ``normal" disequilibrium chemistry model, which assumes that the species is in chemical equilibrium deep in the atmosphere and is quenched at higher altitudes due to vigorous vertical mixing.

\ce{C2H2} is observed in Jupiter \citep{Gladstone1999,Fouchet2000}, where it is formed via photo-chemistry from \ce{CH4} \citep{Tsai2021}, but there is no star to drive photo-chemistry in \w0458, and the brown dwarfs \citep[with mutual separation $5.0^{+0.3}_{-0.6}$\,au,][]{Leggett2019} are not emitting UV flux sufficient to drive this photo-chemistry. Note also that photo-dissociation driven by an external source should lead to \ce{C2H2} high in the upper atmosphere (above $10^{-3}$\,bar), while our MIRI observations are primarily sensitive between 0.1-10\,bar. 

\begin{figure*}[ht]
    \centering
    \includegraphics[width=0.46\textwidth]{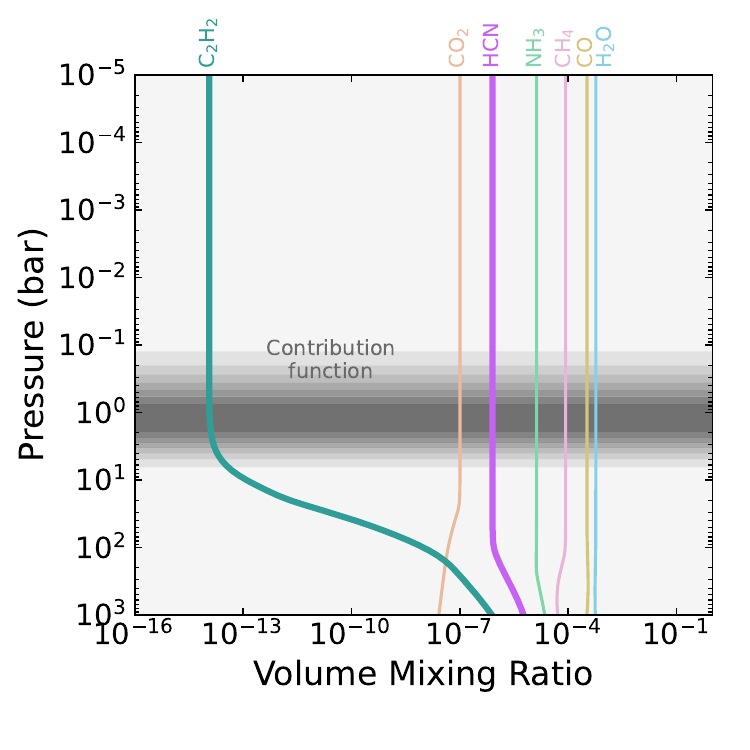}
    \includegraphics[width=0.46\textwidth]{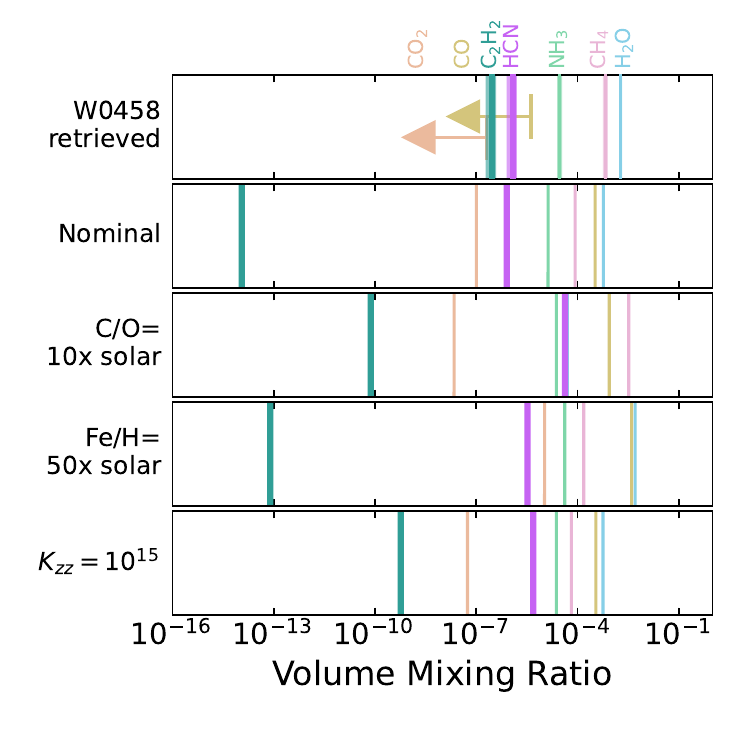}
    \caption{Predictions of atmospheric abundance for varying assumptions about the chemistry and vertical mixing, compared with our retrieved abundances in the \w0458 atmosphere. \textbf{Left:} Atmospheric profiles for a nominal model ($K_{zz}=10^{11}$\,cm$^2$s$^{-1}$, C/O=1$\times$ solar, Fe/H=1$\times$ solar). With strong vertical mixing, we probe only the quenched abundances of these species; \ce{HCN} is under-predicted and \ce{C2H2} is drastically under-predicted in these models relative to our observations. \textbf{Right:} quenched atmospheric abundances, for the nominal model and three models with increased $K_{zz}$, C/O and Fe/H, alongside the retrieval abundances. Even unphysical values for the vertical mixing, C/O or metallicity are insufficient to boost \ce{C2H2} to the observed levels; raising the \ce{C2H2} abundance by increasing C/O or Fe/H also increases \ce{CO} abundance to well above our upper limit. All models are produced with \texttt{VULCAN} \citep{Tsai2017,Tsai2021}.}
    \label{fig:vulcanmodels}
\end{figure*}

We modeled the \w0458 atmospheric chemistry with \texttt{VULCAN} to further explore the conditions under which \ce{C2H2} should be expected in a non-irradiated atmosphere with no thermal inversion. \texttt{VULCAN} is a chemical kinetics code that computes atmospheric composition considering mixing thermochemistry and optionally photochemistry, which has been validated against measurements of Jupiter as well as several exoplanets \citep{Tsai2017,Tsai2021}. The code assumes a one-dimensional model atmosphere, $>$300 chemical reactions (here we use the N-C-H-O chemical network), and uses eddy diffusion to mimic atmospheric dynamics. For this experiment, we used a $P$-$T$ profile from \citet{Linder2019} with $T_\textrm{{eff}}=500$\,K and $\log{g}=5.0$. For our simple toy model we first assumed solar abundances, a vertically constant mixing profile with $K_{zz}=10^{11}$\,cm$^2$s$^{-1}$, and no photo-chemistry. In this model, the expected \ce{C2H2} volume mixing ratio is $\sim10^{-14}$ -- orders of magnitude below our measured value (Fig. \ref{fig:vulcanmodels}, left). We then tested how varying the vertical mixing ($K_{zz}$), the metallicity, or the C/O (via adding carbon atoms, that is, not conserving solar metallicity) impacts predicted \ce{C2H2} values. While increasing any of these parameters does promote \ce{C2H2} production, unphysically high values are required to reach close to the observed value of \ce{C2H2} (Fig.~\ref{fig:vulcanmodels}, right). 

Perhaps, then, there is another un-identified process in the atmosphere. One possibility is that gravity waves are dissipating significant heat in the upper atmosphere, as modeled by \citet{Watkins2010}. Alternatively, perhaps one or both brown dwarfs are magnitically active, driving aurorae and in turn generating UV photons and/or heating the upper atmosphere and driving a temperature inversion. For the brown dwarf WISE J193518.59-154620.3 \citep{Marocco2019}, an auroral process was inferred based on the observation of methane emission features as observed in JWST/NIRSpec G395H spectra \citep{Faherty2024}. However, NIRSpec data for \w0458 does not show the 3.325\,\micron~emission feature that was used in \citet{Faherty2024} to identify aurorae. Galactic cosmic rays have also been theorized to be able to generate \ce{C2H2} in the upper atmosphere \citep{Rimmer2014}, though this would likely only produce smaller quantities of \ce{C2H2} than observed and at higher altitudes than our contribution function indicates. Another possibility is a more exotic scenario, such as a tidally heated exoplanet or exomoon (similar to Io in orbit around Jupiter) providing additional heat to the upper atmosphere. While occurrence rates of small bodies orbiting brown dwarfs are not known, in this case we would expect significant stochasticity in the occurrence and abundance of \ce{C2H2}, based on the presence and exact properties of orbiting small bodies in different systems. Finally, substantial lightning discharges might ionize species in the atmosphere and drive the observed chemistry; a process that has previously been theorized to generate HCN in brown dwarf atmospheres \citep{Helling2019}. 

Even additional energy sources would only be expected to produce \ce{C2H2} in a carbon-rich (C/O$>$1) environment, and not the carbon-poor atmosphere observed here. The detection of \ce{C2H2} remains a challenge for atmospheric models, suggesting a shortcoming in our understanding of vertical transport and disequilibrium processes, in our understanding of the energy sources of this atmosphere, or in our understanding of \ce{C2H2} chemistry \citep[e.g.][]{Saheb2018}.

\subsection{Other hydrocarbons}
\label{sec:c2h4+}

Motivated by the detection of \ce{C2H2}, we also searched for other hydrocarbon species in the \w0458 spectrum, namely \ce{C2H4} and \ce{C2H6}. Our choice of test hydrocarbons is motivated by the detailed chemical study of \citet{Venot2015}, which modeled irradiated atmospheres with and without photo-dissociation. Without photo-dissociation, and for a cool (500K) atmosphere, they predict low volume mixing ratios ($<10^{-7}$) for all carbon species except for \ce{CO}, \ce{CO2}, \ce{CH4} and \ce{CH3}. With photo-dissociation, \ce{HCN}, \ce{C2H2}, \ce{C6H8}, \ce{C2H4} and \ce{C2H6} are predicted to have maximum mixing ratios $>10^{-7}$, and may become relevant for the atmosphere. In each case, the abundance is somewhat higher when C/O is greater than unity. Of course, the thermal profile of these irradiated atmospheres is somewhat different than the expected thermal profile of \w0458, but in the absence of more suitable models that explain the surprising detection of \ce{C2H2}, this work provides the least-worst predictions of relevant hydrocarbon species. We therefore searched for \ce{C2H4} and \ce{C2H6}. There is no evidence for \ce{C2H6}, for which we place an upper limit of $2.34\times10^{-5}$ (at 3$\sigma$). Our model shows tentative evidence for \ce{C2H4} (Bayes factor 1.8, corresponding to a 3.3$\sigma$ detection) at a volume mixing ratio of ${3.0}^{+3.7}_{-2.9}\times10^{-6}$. However, the posterior distribution includes model atmospheres with no \ce{C2H4}, and when visually examining the best-fit spectra with and without \ce{C2H4} we do not see any clear absorption features attributed to \ce{C2H4}, but only subtle changes to the absorption features for other molecules. Without further analysis, it is premature to claim a \ce{C2H4} detection in \w0458, especially before the uncertainties and correlations in the MIRI/MRS data are well understood. We did not search for \ce{C6H8}, which is a remarkably complex molecule to form without an external irradiation source (see \citealt{Venot2015}).

\subsection{Non-detections of \ce{^{15}NH3}, \ce{PH3} and \ce{H2S}}

We tested for \ce{^{15}NH3} in the \w0458~atmosphere. \ce{^{15}NH3} abundance ratios may indicate the formation pathways of brown dwarfs and exoplanets \citep{Nomura2022,Barrado2023}, and \ce{^{15}NH3} has been detected in other cold brown dwarfs. Our retrievals indicate a 3$\sigma$ upper limit of \ce{^{15}NH3} $<$ $2.8\times10^{-6}$, and a ratio $^{14}$N/$^{15}$N $>$ 115. This is consistent with the $^{14}$N/$^{15}$N ratio in cold atmospheres where \ce{^{15}NH3} has been measured: WISE J1828 ($\sim$380\,K) and WISE J0855 ($\sim$285\,K) have $^{14}$\ce{NH3}/\ce{^{15}NH3} ratios of 670$^{+390}_{-211}$ and 349$^{+53}_{-41}$ respectively \citep{Barrado2023,Kuehnle2024}; Jupiter has $^{14}$\ce{NH3}/\ce{^{15}NH3}=435$\pm$50 \citep{Owen2001} and for the ISM $^{14}$N/$^{15}$N=274$\pm$18 \citep{Ritchey2015}. The \w0458 lower limit is consistent with all of these values, meaning that our \ce{^{15}NH3} abundance upper limit does not have the power to differentiate between formation models. 

The detection of the isotopologue \ce{^{15}NH3} is strongly temperature-dependent, as has also been theoretically demonstrated for isotopologues of C- and H- containing molecules \citep{MolliereSnellen2019}. While the absolute flux of warm atmospheres is higher, the absorption features of the main isotopologue species are dominant, and blanket any features from other isotopologues. At cooler temperatures, this blanketing effect is reduced and absorption features of the minor isotopologue can also be observed. We demonstrate this for \ce{NH3} in Appendix \ref{app:15nh3}. \ce{^{15}NH3} will likely only be detectable in planets and brown dwarfs colder than $\sim$500K, though a precise analysis of the trade-off between source flux, \ce{^{14}NH3} abundance, and \ce{^{15}NH3} detectability is beyond the scope of this work.

We tested for \ce{PH3} in the \w0458~atmosphere, and constrain the volume mixing ratio to $<2.5\times10^{-6}$ at 3$\sigma$. This is consistent with other early JWST observations of cold brown dwarfs \citep{Beiler2023,Beiler2024b}. While the species is known in the atmospheres of Jupiter and Saturn \citep{Prinn1975,Ridgway1976,Gillett1974}, phosphine has not yet been detected in any exoplanet or brown dwarf except very recently in WISE-0855 \citep{Rowland2024} at a volume mixing ratio of $(5.7^{+1.0}_{-0.9})\times10^{-10}$ or $(1.23\pm0.2)\times10^{-9}$ depending on the dataset used, and possibly in UNCOVER-BD-3 \citep{Burgasser2024} though this is disputed \citep{Beiler2024b}. NIRSpec observations in the 4-5\,\micron~range are more sensitive to \ce{PH3} than our data, though a sufficiently high \ce{PH3} abundance would also be visible in MIRI/MRS spectra at 8.9\,\micron~and 10.1\,\micron.

Finally, we tested for the presence of \ce{H2S}, which has been seen in a few other cold brown dwarfs with JWST \citep{Hood2024,Lew2024}, but we do not detect the species and place a modest upper limit of $1.8\times10^{-3}$. This is unsurprising without NIRSpec data, since \ce{H2S} has no strong features beyond 5\,\micron.

\section{Discussion}
\label{sec:discussion}

\begin{figure}[t]
    \centering
    \includegraphics[width=0.48\textwidth]{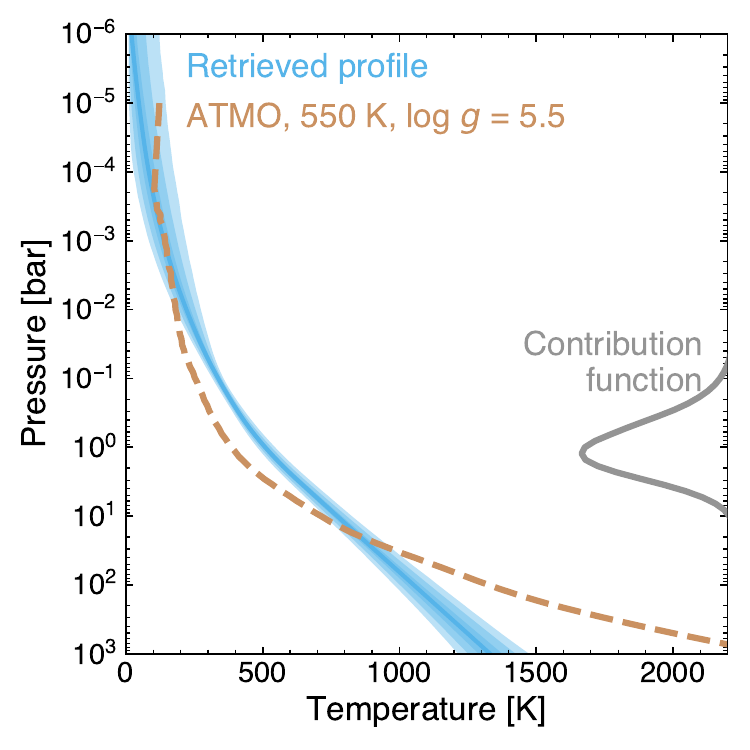}
    \caption{Retrieved pressure-temperature profile of the two brown dwarfs (blue), with the 1$\sigma$, 2$\sigma$ and 3$\sigma$ intervals shaded, and the contribution function (grey). The $P$-$T$ profile is broadly consistent with a self-consistent profile (here an ATMO model (brown) with 550\,K, $\log{g}$=5.5, and chemical equilibrium \citep{Phillips2020} for reference). The retrieved profile has slightly less curvature -- which is unsurprising, given that we use a regularization term for the $P$-$T$ profile (see text).}
    \label{fig:ptprofile}
\end{figure}

The mid-IR spectrum of \w0458 is well fitted with a clear atmosphere model, and the retrieved temperature-pressure profile is broadly consistent with predictions from self-consistent forward-models (see Fig.~\ref{fig:ptprofile}), indicating that our $P$-$T$ parametrization within the \texttt{pRT} setup is sufficient to describe this atmosphere. Our model shows evidence for five molecules: \ce{H2O}, \ce{CH4}, \ce{NH3}, \ce{HCN} and \ce{C2H2}.

We measure a radius of $0.807^{+0.008}_{-0.007}\Rjup$ for each brown dwarf (assuming both binary components have equal radii, and based on the effective radius of $1.141^{+0.011}_{-0.010}\Rjup$ determined for the pair in our retrievals), while the ATMO evolutionary models predict a radius of 0.84\Rjup~for a single field-age, 40\Mjup~brown dwarf and 0.93\Rjup~for a single field-age, 20\Mjup~brown dwarf. The measured uncertainty on the effective radius is very small (0.01\Rjup), while the difference in effective radius between our two models (Table \ref{tab:retrievals}) is 0.05\Rjup: this highlights the model-dependency of the radius, which is also degenerate with the measured temperature and molecular abundances. Further, the distance is fixed to 9.24pc in our calculations, and the distance uncertainty ($\pm0.15$\,pc) has not been incorporated into the radius uncertainty. Taking this into account, the retrieved radius is broadly consistent with predictions of evolutionary models.

We derive a mass of $66^{+19}_{-15}\Mjup$ for each brown dwarf based on the retrieved values of $\log{g}$ and radius values, corresponding to a total mass of $132^{+38}_{-28}\Mjup$ for the pair. This is higher than expected from evolutionary models, based on the cold temperature of the pair. The retrieved mass is also somewhat higher than the dynamical mass estimated for the pair of both brown dwarfs, ($70^{+15}_{-24}\Mjup$ \citealt{Leggett2019}), though we note that using a single $\log{g}$ value in our retrievals for the binary is an over-simplification. Future works should use the resolved brown dwarf astrometry from MIRI and/or NIRSpec to improve this orbital fit, and make a more robust comparison between the spectroscopically-derived and dynamically-derived masses. 

Analysis of the NIRSpec observations of this target (Lew et al.~in prep) will likely improve our CO and CO$_2$ constraints, and help determine a more precise C/O ratio for the atmosphere. This may shed further light on the carbon-chemical network of \w0458, which will be relevant for explaining the presence of \ce{C2H2}. We do not expect this to significantly alter the constraints, since \ce{H2O} and \ce{CH4} are the main carriers of oxygen and carbon respectively in a cool, high-$\log{g}$ atmosphere \citep[e.g.][]{Zahnle2014}, and our \ce{CO} and \ce{CO2} upper limits are much lower than the measured \ce{CH4} abundance suggesting that \ce{CH4} is a sufficient proxy for atmospheric carbon content. Note also that sufficiently high \ce{CO2} abundance would lead to an absorption feature at 15\,\micron, which would be readily detectable with MIRI/MRS data but is not seen in the \w0458 spectrum. If there is any sequestration of oxygen in cloud species, then this would lead to an even lower C/O ratio than our measured value (and not the strongly super-solar C/O regime that might be needed to promote \ce{C2H2} formation in current models).

\subsection{Uncertainty Inflation}

We inflated the pipeline-derived flux uncertainties, as discussed in Section \ref{sec:retrievals}. For the model including \ce{HCN} and \ce{C2H2}, we retrieved a \textit{b}-parameter value of $-8.208\pm0.017$, corresponding to an uncertainty inflation factor of $19\pm$4 (full range 9-34) depending on the wavelength. In fact, this relatively high inflation is a natural consequence of a known issue in the JWST data reduction: the uncertainty calculated by the pipeline is underestimated\footnote{see issue MR-MRS06 of \url{https://jwst-docs.stsci.edu/known-issues-with-jwst-data/miri-known-issues/miri-mrs-known-issues}}; this underestimation can be up to 10-50$\times$. The proposed workaround is to bootstrap errors based on the science spectra; in our analysis the inclusion of a \textit{b}-parameter inflating the uncertainties is functionally equivalent to this suggested bootstrapping. Taking this known issue into account, the uncertainty inflation is reasonable, indicating that the atmospheric structure and chemical composition of the brown dwarf atmosphere of \w0458 is relatively well-captured in our best-fit model. An ideal solution to the underestimated errors would also reflect any wavelength dependence of this uncertainty (whereas our \textit{b}-parameter applies the same scaling factor, added in quadrature, across all wavelengths). However, this would introduce additional free parameters; the residuals (panels c, e, g of Fig.~\ref{fig:mainspectrum_and_model}) do not show any strong wavelength-dependence, indicating that the \textit{b}-parameter approach is sufficient to model the error underestimation in this case. This assumption should be revisited if a future pipeline update improves the error estimation.

\subsection{Optimal wavelengths for \ce{HCN} and \ce{C2H2} detection}

\ce{HCN} and \ce{C2H2} are most detectable at mid-IR wavelengths, with the JWST/MIRI MRS instrument. In Figure \ref{fig:hcnopacity} we present opacities of the molecular species in our best-fit model, based on the retrieved abundance value or upper limit from our best-fit model; \ce{HCN} and \ce{C2H2} are highlighted with bold lines in this plot. For this figure we used a nominal temperature of 600\,K and pressure of 0.1\,bar. In the lower panel, we highlight regions of the NIRSpec (left) and MRS (right) wavelength ranges where \ce{HCN} and \ce{C2H2} have the highest opacities. At the abundances observed in \w0458, the only significant absorption features of \ce{HCN} are at 14\,\micron~and 14.3\,\micron, and the only significant feature of \ce{C2H2} is at 13.7\,\micron. Both species also have a significant impact on the continuum, especially between $\sim$12-16\,\micron. 

Considering the 1-5\,\micron wavelength range (i.e., the wavelengths observable with JWST/NIRSpec), neither \ce{HCN} nor \ce{C2H2} is likely to be detectable at the abundance measured in \w0458: for an R$\sim$1,000 spectrum and the \w0458 abundances, \ce{HCN} is the strongest absorber only for a single data point at 3.06\,\micron, where the \ce{H2O} opacity briefly dips and before the \ce{CH4} opacity becomes strong. For a sufficiently elevated abundance, \ce{HCN} may become visible around 3\,\micron~and possibly 1.5\,\micron~in NIRSpec observations. \ce{C2H2} has an even lower opacity in this region, and we do not expect \ce{C2H2} to be observable in the near-IR. Further, neither species should be detectable with MIRI/LRS, which is optimized for sensitivity between $\sim$5-12\,\micron~\citep{Kendrew2015}. The MIRI/MRS instrument is therefore uniquely sensitive to low abundance detections of \ce{HCN}  and \ce{C2H2} in brown dwarf atmospheres, and subchannel 3B is required for both molecules.

\subsection{Future prospects for \ce{HCN} and \ce{C2H2} detection}

\w0458~is the first T-type brown dwarf observed with MIRI/MRS, and as such the first T-type brown dwarf where \ce{HCN} and \ce{C2H2} are readily detectable. Two Y-type objects have MIRI/MRS spectra \citep{Barrado2023,Kuehnle2024}, but \ce{HCN} and \ce{C2H2} are not detected in either object.

It remains to be seen how common these species are in other T-type atmospheres. In particular, in-hand observations of the T8 object Ross-458 (ExoMIRI consortium, Whiteford et al.~in prep) and upcoming observations of the brown dwarf binary Eps Ind BA/BB (T1+T6, \citealt{King2010,Matthews_jwstgo5765}) will provide a valuable comparison of how \ce{HCN} and \ce{C2H2} abundances vary with temperature. The T-type exoplanets COCONUTS-2b (T9, \citealt{Kirkpatrick2011,Zhang2021}) and GJ504b (T8, \citealt{Janson2013}) are also scheduled for MIRI/MRS observations \citep{Patapis_jwstgo3647,Patapis_jwstgo6463}, though these have much lower surface gravity and lower \ce{HCN} abundances are expected in the low $\log{g}$ regime \citep{Zahnle2014}. Non-detections of \ce{HCN} in these low-surface-gravity objects would provide further indications that \ce{HCN} is indeed driven by strong vertical mixing, as modeled in \citet{Zahnle2014}. The detection or absence of \ce{C2H2} as a function of effective temperature and surface gravity may explain why we are detecting such a surprisingly high \ce{C2H2} abundance in \w0458.

\section{Conclusions}
\label{sec:conclusions}

In this work, we have presented a mid-IR spectrum of \w0458 at resolution $>$1000 between 4.9\,\micron~and 18.0\,\micron, and treated the system as an unresolved binary. This is the first JWST/MIRI/MRS spectrum of any T-type object, and is well-modeled by a clear, molecule-rich atmosphere in our retrieval analysis. 
Our key findings are the following:

   \begin{enumerate}
      \item \ce{HCN} is detected at 13.4$\sigma$ in the \w0458 atmosphere, suggesting very strong vertical mixing ($K_{zz}\sim10^{11}$\,cm$^2$s$^{-1}$), in tension with previous studies of T-dwarf vertical mixing.
      \item \ce{C2H2} is detected at 9.5$\sigma$ in the \w0458 atmosphere; this discovery is surprising and cannot be explained with a standard prescription of quenching, i.e., a deep-atmosphere equilibrium species brought to higher altitudes (lower pressure levels) with vertical mixing using a single $K_{zz}$ value.
      \item \ce{PH3} and \ce{^{15}NH3} are not detected. The \ce{PH3} non-detection is consistent with other JWST spectra of late-type brown dwarfs. The \ce{^{15}NH3} non-detection is consistent with the observed $^{15}$N enrichment of other cold brown dwarfs, and we highlight that \ce{^{15}NH3} will be most readily detectable in cold brown dwarfs ($T_\textrm{eff} \lesssim 500$\,K). 
      \item The atmosphere of \w0458 is well-fitted by our retrievals, with a cloud-free model. The required uncertainty inflation is consistent with the known underestimation of MRS uncertainties in the \texttt{jwst} pipeline.
   \end{enumerate}

This spectrum demonstrates the power of MIRI MRS to characterize cold brown dwarfs. Future works should study the \ce{HCN} and \ce{C2H2} in more detail, and determine whether these species are present in other cold brown dwarfs of similar temperature to \w0458. Our data challenge existing prescriptions for atmospheric mixing and/or \ce{C2H2} chemistry, and hint towards a new era in brown dwarf chemistry with an increasing number of carbon species detectable in these cold atmospheres.

\vspace{0.5cm}

\begin{footnotesize}

\noindent
This work is based on observations made with the NASA/ESA/CSA James Webb Space Telescope. The data were obtained from the Mikulski Archive for Space Telescopes at the Space Telescope Science Institute, which is operated by the Association of Universities for Research in Astronomy, Inc., under NASA contract NAS 5-03127 for JWST. These observations are associated with program \#1189. All the {\it JWST} data used in this paper can be found in MAST: \dataset[10.17909/0tbz-1635]{http://dx.doi.org/10.17909/0tbz-1635}.

We used \texttt{pyMultiNest} \citep{Buchner2014}, which in turn builds on \texttt{MultiNest} \citep{Feroz2008,Feroz2009,Feroz2019} for the nested sampling analysis.  We used various functions from \texttt{astropy} \citep{astropy_i,astropy_ii,astropy_iii}, \texttt{pandas} \citep{pandas_i,pandas_ii}, \texttt{scipy} \citep{scipy}, \texttt{matplotlib} \citep{matplotlib} and \texttt{seaborn} \citep{seaborn} to carry out analysis and create figures.

P.P. thanks the Swiss National Science Foundation (SNSF) for financial support under grant number 200020\_200399.
N.W. acknowledges support from NSF awards \#2238468 and \#1909776, and NASA Award \#80NSSC22K0142
P.-O.L acknowledges funding support by CNES.
S.-M.T. acknowledge support from NASA through Exobiology Grant No. 80NSSC20K1437.
B.V., O.A, I.A, and P.R. thank the European Space Agency (ESA) and the Belgian Federal Science Policy Office (BELSPO) for their support in the framework of the PRODEX Programme.
O.A. is a Senior Research Associate of the Fonds de la Recherche Scientifique – FNRS. 
D.B. is supported by Spanish MCIN/AEI/10.13039/501100011033
grants PID2019-107061GB-C61 and PID2023-150468NB-I00.
G.O. acknowledges support from the Swedish National Space Board and the Knut and Alice Wallenberg Foundation.
J.P.P. acknowledges financial support from the UK Science and Technology Facilities Council, and the UK Space Agency.
E.v.D. acknowledges support from A-ERC grant 101019751 MOLDISK.
T.P.R. acknowledges support from the ERC grant 743029 EASY. 
G.\"O. acknowledges support from SNSA.

\end{footnotesize}

\bibliography{w0458}{}
\bibliographystyle{aasjournal}

\appendix

\section{Posteriors of best-fit Retrievals}
\label{app:posteriors_table_corner}

Here we present the full posteriors of the retrieval analysis, as described in Section \ref{sec:retrievals}. In Table \ref{tab:retrievals} we provide posterior values (median and 1$\sigma$ confidence intervals) for all parameters, for our ``basic" retrieval (molecules \ce{CH4}, CO$_2$, CO, \ce{H2O} and \ce{NH3}) and our retrieval with \ce{HCN} and \ce{C2H2} (and otherwise identical to the ``basic" retrieval). 

\begin{figure*}
    \centering
    \includegraphics[width=1.\textwidth]{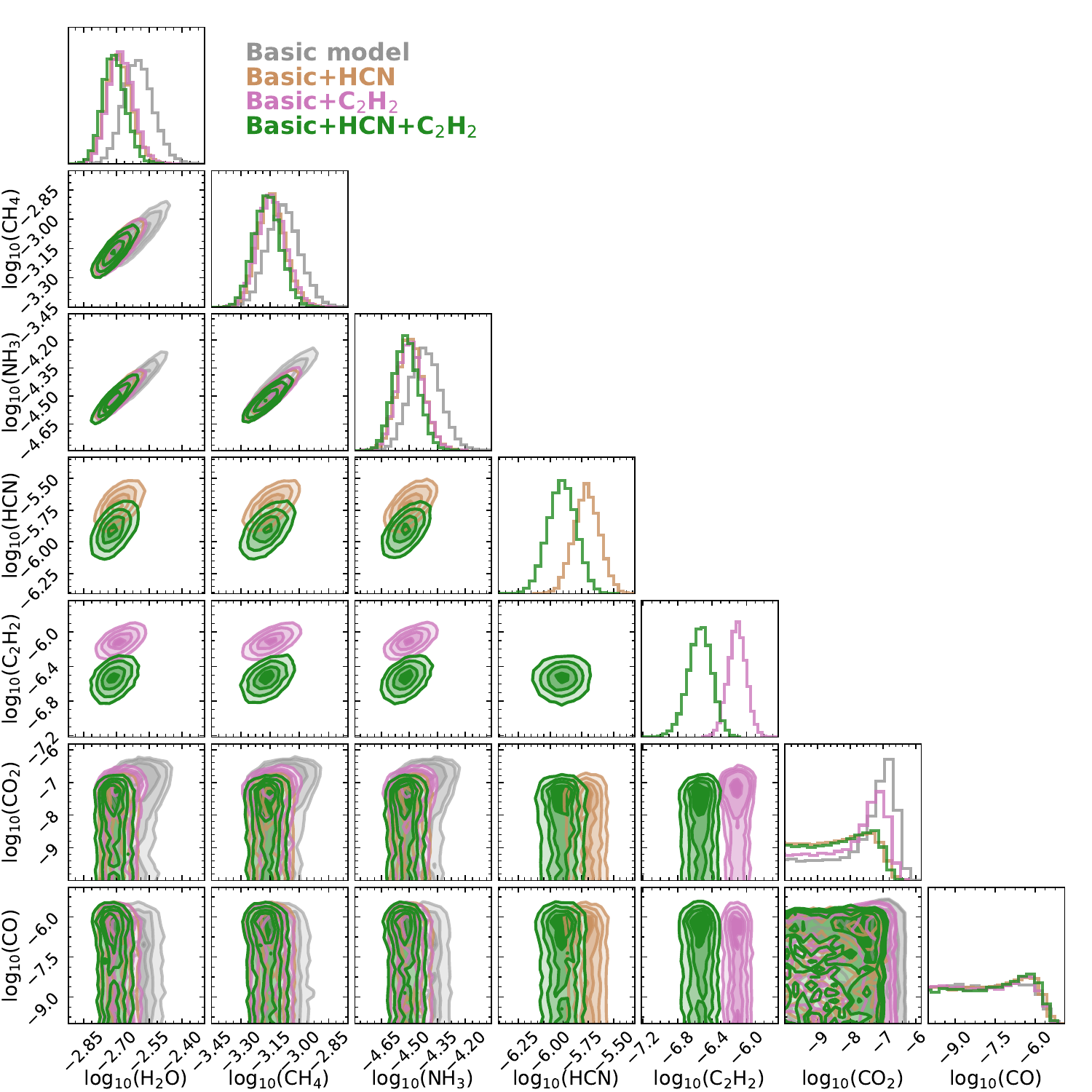}
    \caption{Corner plot for molecular abundances in the \texttt{pRT} retrievals. Abundances shown are volume mixing ratios. We show the posteriors for our basic model (grey, incl.~\ce{H2O}, \ce{CH4}, \ce{NH3}, CO, CO$_2$) as well as with \ce{HCN} added (brown), with \ce{C2H2} added (pink), and with both \ce{HCN} and \ce{C2H2} added (green). \ce{HCN} and \ce{C2H2} have some overlapping features around 14\,\micron, and have similar impacts on the continuum (see Figure \ref{fig:hcnopacity}; this causes the abundances to be correlated and the inclusion of one species alters the retrieved abundance of the other. For the models without \ce{HCN}, a spurious tentative detection of CO$_2$ is present in the posteriors, as discussed in the text; this disappears when \ce{HCN} is included in the models. The retrieved abundances of \ce{H2O}, \ce{CH4} and \ce{NH3}, all of which have strong features in the MRS spectrum, are largely unchanged by the addition of \ce{HCN} and \ce{C2H2}.}
    \label{fig:prt_corner_abundance}
\end{figure*}

In Figure \ref{fig:prt_corner_abundance} we provide a corner plot for the retrieved volume mixing ratios for all molecules in the atmospheres, for our ``basic" model and for models with \ce{HCN} added, and with \ce{C2H2} added, and with both \ce{HCN} and \ce{C2H2} added. The abundances of \ce{HCN} and \ce{C2H2} are somewhat correlated, since both have a similar effect on the continuum between 12-16\,\micron. The basic model and the model with \ce{C2H2} added show a hint of a CO$_2$ detection, in the form of a peak in the posterior at a volume mixing ratio of $\sim10^{-7}$ but a long tail towards lower abundances. However, this posterior peak disappears when including \ce{HCN}; we infer that the \ce{CO2} detection was a false positive to explain the broadband spectral shape effect of \ce{HCN}. 

\renewcommand{\arraystretch}{1.3}
\begin{table*}
    \caption{Best-fit parameters for our retrievals. We show the values for both our ``basic model" (with 5 molecules: \ce{H2O}, \ce{CH4}, \ce{NH3}, \ce{CO} and \ce{CO2}), as well as our preferred model which is the same as the basic model except for the addition of \ce{HCN} and \ce{C2H2}. Listed uncertainties are 1$\sigma$.}       
    \label{tab:retrievals}
    \centering
    \begin{tabular}{c|c|c|c}
\hline
\hline
\textbf{Parameter} & \textbf{Prior} & \textbf{Basic model} & \textbf{Model w/ \ce{HCN}, \ce{C2H2}} \\ 
\hline
\hline
\multicolumn{4}{l}{\textit{Physical Parameters}} \\
\hline
 log(g[cm/s$^{-2}$]) & Uniform [2.5, 6.0] & $5.56^{+0.13}_{-0.13}$ & $5.40^{+0.11}_{-0.12}$ \\
 Measured Radius [$R_\textrm{{Jup}}$] & Uniform [0.46, 2.74] & $1.094^{+0.010}_{-0.010}$ & $1.141^{+0.011}_{-0.010}$ \\
\hline
\multicolumn{4}{l}{\textit{Pressure-Temperature Profile}} \\
\hline
 Base Temperature (Node0) [K] & Uniform [100, 9000] & $1334^{+43}_{-41}$ & $1335^{+45}_{-42}$ \\
 Temp.~Ratio Node1/Node0 & Uniform [0.2, 1.0] & $0.79^{+0.01}_{-0.01}$ & $0.79^{+0.01}_{-0.01}$ \\
 Temp.~Ratio Node2/Node1 & Uniform [0.2, 1.0] & $0.74^{+0.01}_{-0.01}$ & $0.73^{+0.01}_{-0.01}$ \\
 Temp.~Ratio Node3/Node2 & Uniform [0.2, 1.0] & $0.66^{+0.01}_{-0.01}$ & $0.66^{+0.01}_{-0.01}$ \\
 Temp.~Ratio Node4/Node3 & Uniform [0.2, 1.0] & $0.65^{+0.01}_{-0.01}$ & $0.66^{+0.02}_{-0.01}$ \\
 Temp.~Ratio Node5/Node4 & Uniform [0.2, 1.0] & $0.63^{+0.03}_{-0.03}$ & $0.66^{+0.04}_{-0.03}$ \\
 Temp.~Ratio Node6/Node5 & Uniform [0.2, 1.0] & $0.59^{+0.06}_{-0.06}$ & $0.63^{+0.06}_{-0.06}$ \\
 Temp.~Ratio Node7/Node6 & Uniform [0.2, 1.0] & $0.57^{+0.08}_{-0.09}$ & $0.60^{+0.08}_{-0.08}$ \\
 Temp.~Ratio Node8/Node7 & Uniform [0.2, 1.0] & $0.55^{+0.10}_{-0.09}$ & $0.57^{+0.10}_{-0.09}$ \\
 Temp.~Ratio Node9/Node8 & Uniform [0.2, 1.0] & $0.50^{+0.24}_{-0.17}$ & $0.47^{+0.23}_{-0.16}$ \\
 $\gamma$ & Gaussian, $\sigma$=1, $\mu$=1 & $15^{+4}_{-3}$ & $17^{+5}_{-3}$ \\
\hline
\multicolumn{4}{l}{\textit{Molecular Abundances [Volume Mixing Ratio]}} \\
\hline
 \ce{H2O} & Uniform in log(mass fraction) between [-10,0] & $2.5^{+0.4}_{-0.4} \times 10^{-3}$ & $1.9^{+0.3}_{-0.2} \times 10^{-3}$ \\
 \ce{CH4} & Uniform in log(mass fraction) between [-10,0] & $8.3^{+1.8}_{-1.4} \times 10^{-4}$ & $6.8^{+1.2}_{-1.0} \times 10^{-4}$ \\
 \ce{NH3} & Uniform in log(mass fraction) between [-10,0] & $3.8^{+0.8}_{-0.6} \times 10^{-5}$ & $3.0^{+0.5}_{-0.4} \times 10^{-5}$ \\
 \ce{HCN} & Uniform in log(mass fraction) between [-10,0] & -- & $1.2^{+0.4}_{-0.3} \times 10^{-6}$ \\
 \ce{C2H2} & Uniform in log(mass fraction) between [-10,0] & -- & $2.9^{+1.0}_{-0.8} \times 10^{-7}$ \\
\hline
\multicolumn{4}{l}{\textit{Abundance Upper Limits [Volume Mixing Ratio]}} \\
\hline
 \ce{CO} & Uniform in log(mass fraction) between [-10,0] & $< 4.1 \times 10^{-6}$ & $< 4.2 \times 10^{-6}$ \\
 \ce{CO2} & Uniform in log(mass fraction) between [-10,0] & $< 5.8 \times 10^{-7}$ & $< 2.0 \times 10^{-7}$ \\
\hline
\multicolumn{4}{l}{\textit{Other Parameters}} \\
\hline
b & Uniform [-14.259, -7.150] & $-8.15^{+0.02}_{-0.02}$ & $-8.21^{+0.02}_{-0.02}$ \\
\hline
\hline
\multicolumn{4}{l}{\textit{Derived Parameters}} \\
\hline
  $T_\textrm{{eff}}$ [K] & -- & $580.3^{+5.0}_{-5.1}$ & $565.7^{+6.5}_{-5.5}$ \\
 C/O [atomic ratio] & -- & $0.33^{+0.03}_{-0.03}$ & $0.35^{+0.03}_{-0.03}$ \\
 Fe/H [$\times$ solar] & -- & $1.71^{+0.31}_{-0.25}$ & $1.35^{+0.19}_{-0.15}$ \\
 Radius/sqrt(2) [$R_\textrm{{Jup}}$] & -- & $0.774^{+0.007}_{-0.007}$ & $0.807^{+0.008}_{-0.007}$ \\
 Mass [$M_\textrm{{Jup}}$] & -- & $88^{+31}_{-23}$ & $66^{+19}_{-15}$ \\
\hline
\hline
\multicolumn{4}{l}{\textit{Fixed Parameters}} \\
\hline
Distance [pc] & -- & 9.24 & 9.24 \\
\hline
\hline
    \end{tabular}
\end{table*}

\section{Detectability of \ce{^{15}NH3}}
\label{app:15nh3}

\ce{^{15}NH3} is not observed in the \w0458 atmosphere, even though it has been seen in other MRS observations of cold brown dwarfs. For the cold brown dwarf WISE J1828 (380K) \ce{^{15}NH3} was detected \citep{Barrado2023}, and for the coldest known brown dwarf WISE J0855 ($\sim$285K), the \ce{^{15}NH3}/$^{14}$\ce{NH3} is very well constrained \citep{Kuehnle2024}. We infer this is a temperature-dependent effect (similar to that described for C- and H-isotopologues in \citealt{MolliereSnellen2019}).

\begin{figure*}
    \centering
    \includegraphics[width=0.99\textwidth]{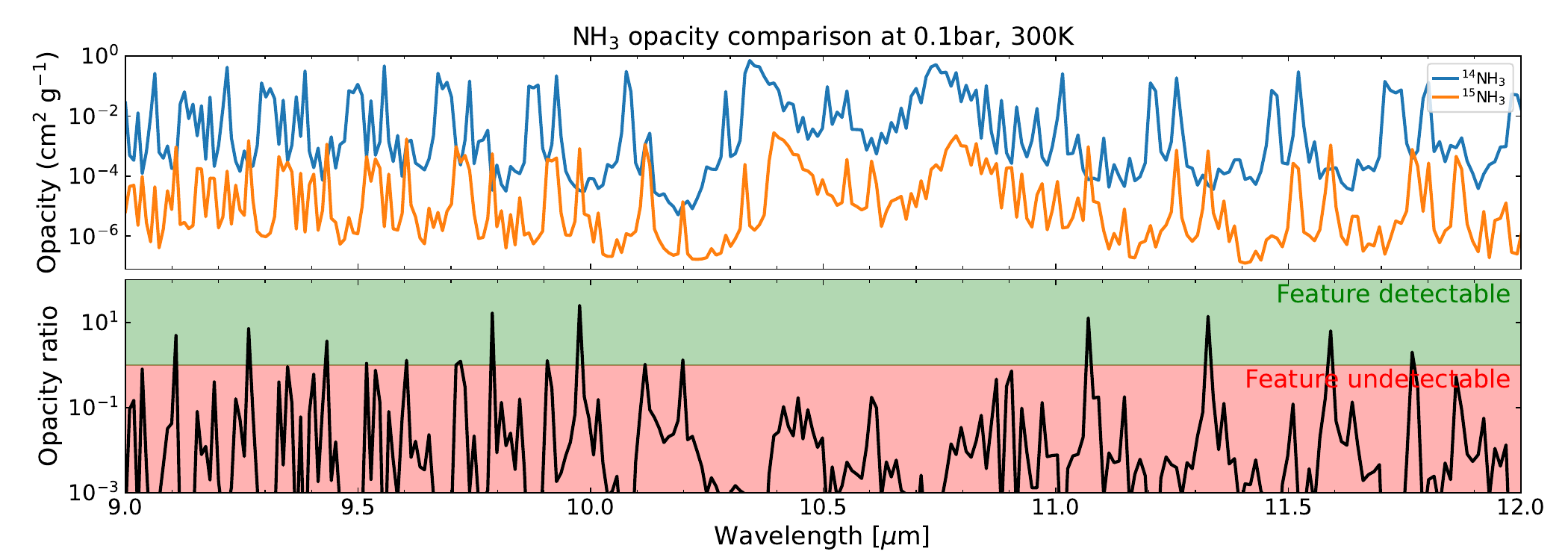}
    \includegraphics[width=0.99\textwidth]{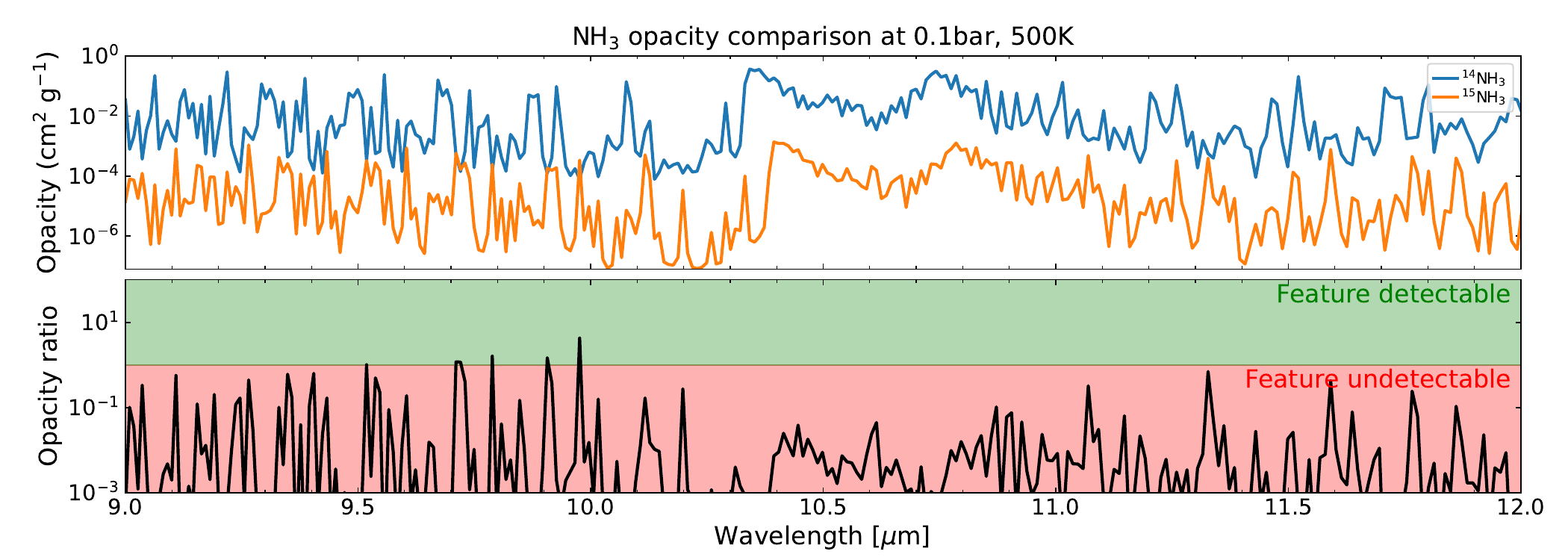}
    \includegraphics[width=0.99\textwidth]{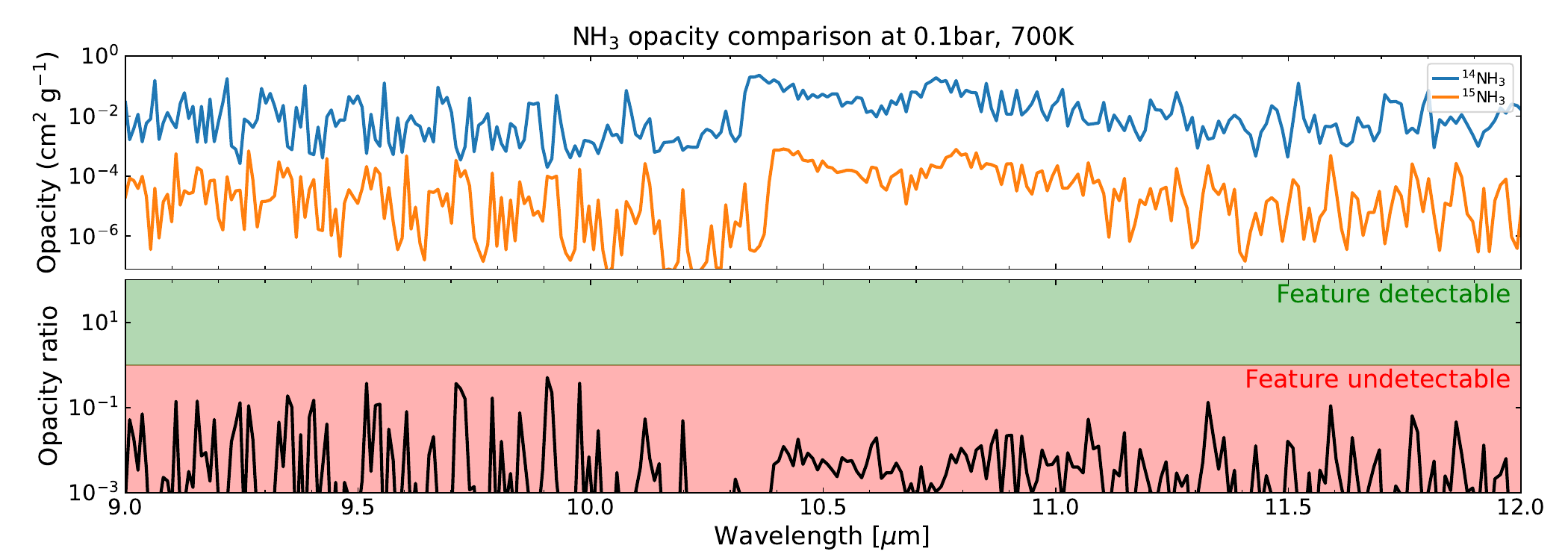}
    \caption{Relative opacities for $^{14}$\ce{NH3} and \ce{^{15}NH3} at three nominal temperatures, assuming an abundance ratio of 250:1 (similar to the ISM ratio for $^{14}$N/$^{15}$N, \citealt{Ritchey2015}), and the ratio between the two opacities in each case. A simplified assumption is that \ce{^{15}NH3} is detectable if it has any features with local opacity stronger than the \ce{^{14}NH3} opacity, i.e., in the green region of the lower panel for each temperature. \ce{^{15}NH3} is only detectable in the coldest brown dwarfs.}
    \label{fig:15nh3opacities}
\end{figure*}

In this appendix we provide a comparison of the \ce{^{15}NH3} and $^{14}$\ce{NH3} opacities, at a reference pressure of 0.1bar and three reference temperatures of 300K, 500K and 700K, and generated with \texttt{pRT} (Figure \ref{fig:15nh3opacities}). For this comparison we assume an abundance ratio of 250, similar to the ISM value. At warm temperatures, the $^{14}$\ce{NH3} opacity is much higher than the \ce{^{15}NH3} opacity at all wavelengths, and \ce{^{15}NH3} is not detectable. At colder temperatures, select \ce{^{15}NH3} opacity features begin to protrude over the $^{14}$\ce{NH3} opacity, meaning that a sufficiently high-quality mid-IR spectrum would detect both isotopes. For sufficiently cold planets (e.g. the 300K example in Figure \ref{fig:15nh3opacities}), many \ce{^{15}NH3} features clearly emerge over the $^{14}$\ce{NH3} opacity. These plots highlight the strong temperature dependence of \ce{^{15}NH3} detectability, which will likely only be detectable in planets and brown dwarfs colder than $\sim$500K.

\section{Custom Background Treatment}
\label{app:backgrounds}

After dither subtraction, the background remains poorly corrected: there is significant spatial structure, primarily in the form of ``striping'', i.e., linear regions across the detector that are significantly non-zero. Even more concerning, this poorly corrected background varies as a function of wavelength, meaning that wavelength-dependent systematics introduce spurious ``features" into the spectrum, or change the shape of true molecular features.  Given the 2D nature of this structure, an annulus subtraction is insufficient to correct the background contribution to extracted photometry. This in turn adds a systematic to the 1D spectrum: aperture photometry on each slice of the 3D cube is biased since the local background contribution is not zero within the aperture, and instead contributes a wavelength-dependent positive or negative flux to the photometry procedure. When the MRS cube is built in the \texttt{ifualign} orientation (i.e., in the instrument plane, as opposed to the \texttt{skyalign}, where the cube is built with North up and East left), these background variations appear as horizontal and vertical stripes in each slice of the 3D cube. This suggests that they are associated with large-scale detector systematics, with pixels that are physically close in the detector space showing correlated values.

\begin{figure}
    \centering
    \includegraphics[width=0.48\textwidth, trim=0cm 0cm 0cm 2cm, clip]{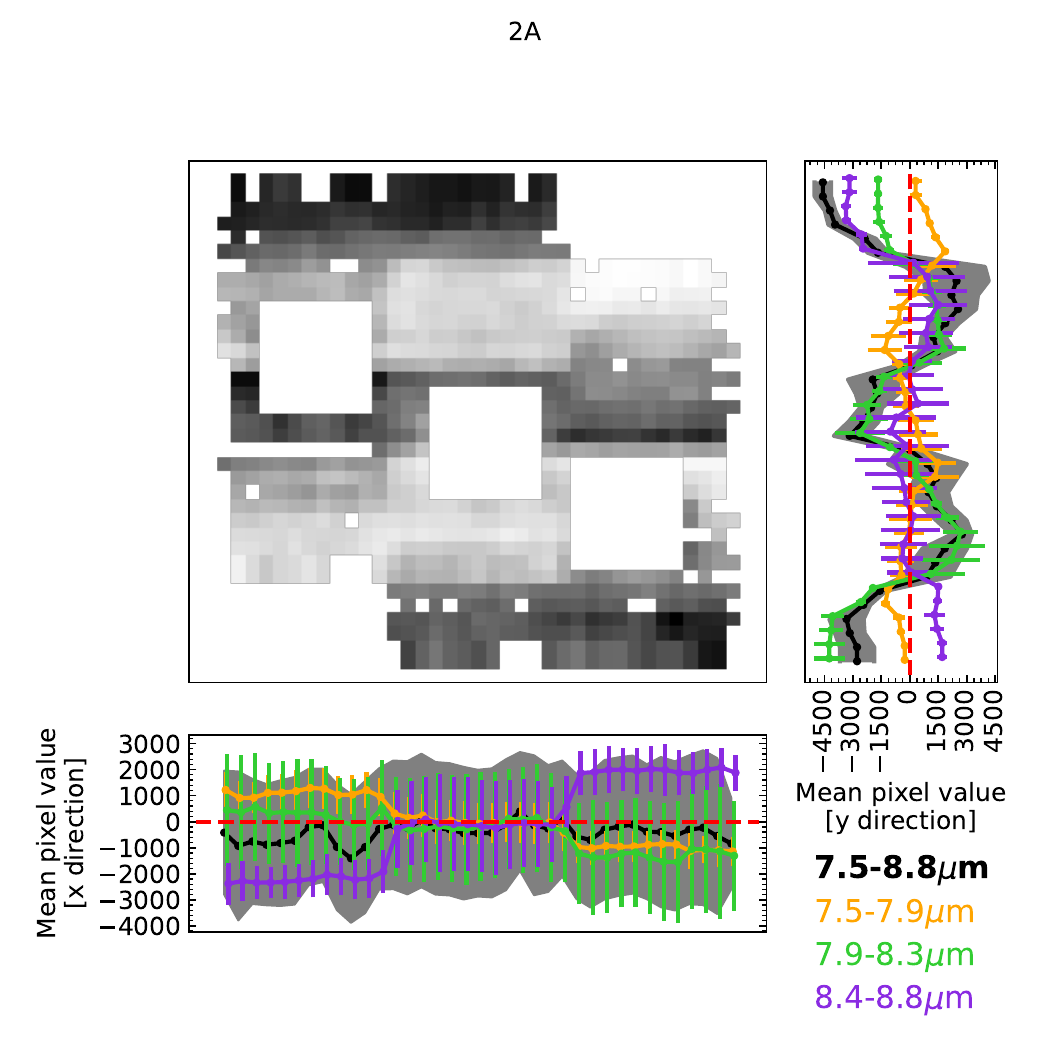}
    \caption{Background systematics in the MIRI/MRS data. The central image shows the channel 2A IFU cube (\texttt{*s3d.fits)} collapsed along the wavelength direction, with the source location and the two negative copies of the source (due to the dither strategy of these observations) masked. Even after dither subtraction, the background of these frames is significantly divergent from 0, and varies with wavelength. In the right hand side and lower panels, black lines and grey shading indicate the mean and 1-$\sigma$ variation of the pixel value along each row/column; orange, green and purple traces indicate the mean and 1-$\sigma$ values for each 1/3 of the wavelength range of the data. These traces show strong variation in the ``normalized" background through the cube, which introduce systematics into the spectrum, especially for faint targets such as \w0458.}
    \label{fig:backgroundvariance}
\end{figure}

Figure \ref{fig:backgroundvariance} illustrates this issue: here we show an example IFU cubes  (\texttt{*s3d.fits}), extracted in the \texttt{ifualign} orientation and collapsed across wavelength and with the source masked. Significant striping in the background is clear, especially in the horizontal direction. The right and lower panels show the mean and variation of the pixels in each row/column, with colored traces demonstrating how this background evolves with wavelength through as single IFU cube.

\begin{figure*}
    \centering
    \includegraphics[width=0.8\textwidth, trim=1cm 0cm 2.3cm 1.1cm, clip]{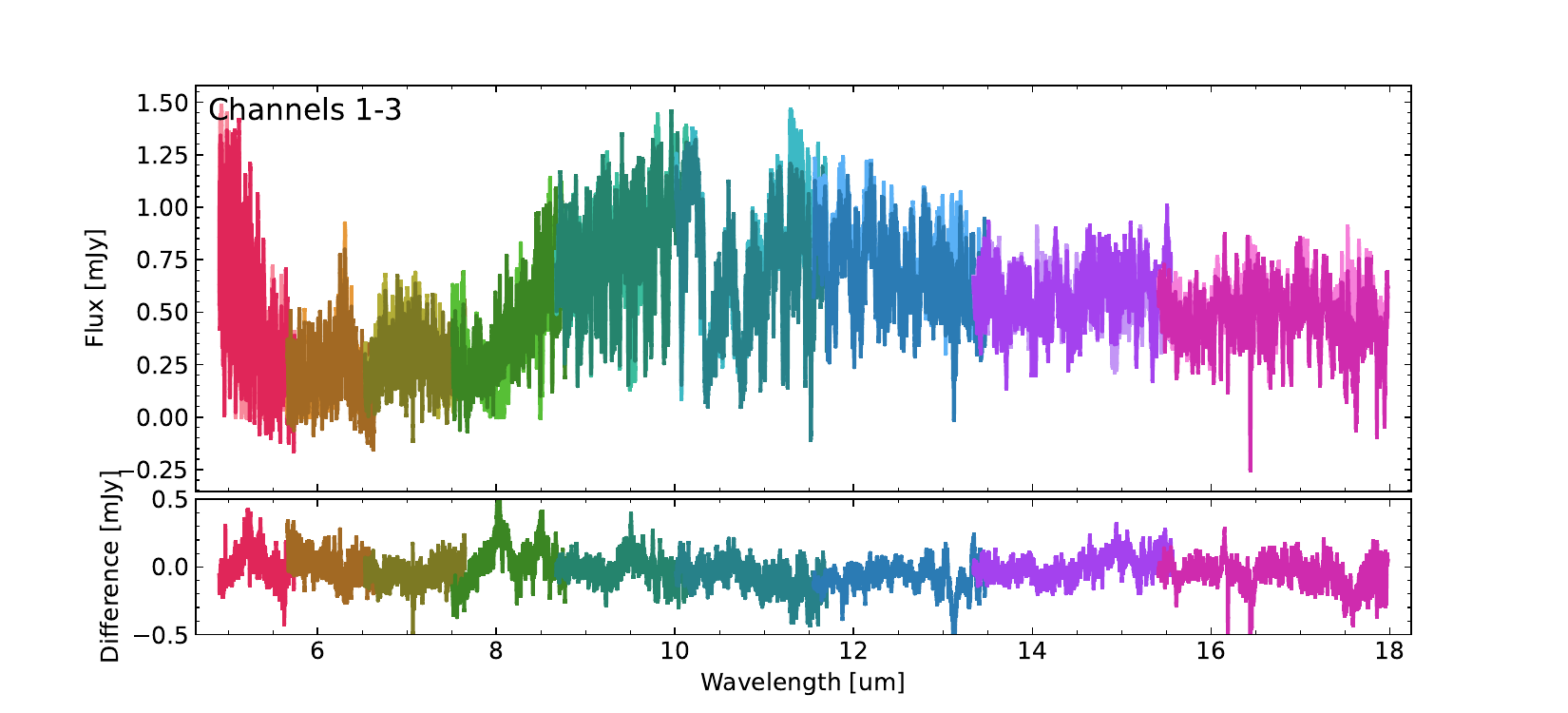}
    \includegraphics[width=0.8\textwidth, trim=1cm 0cm 2.3cm 1.1cm, clip]{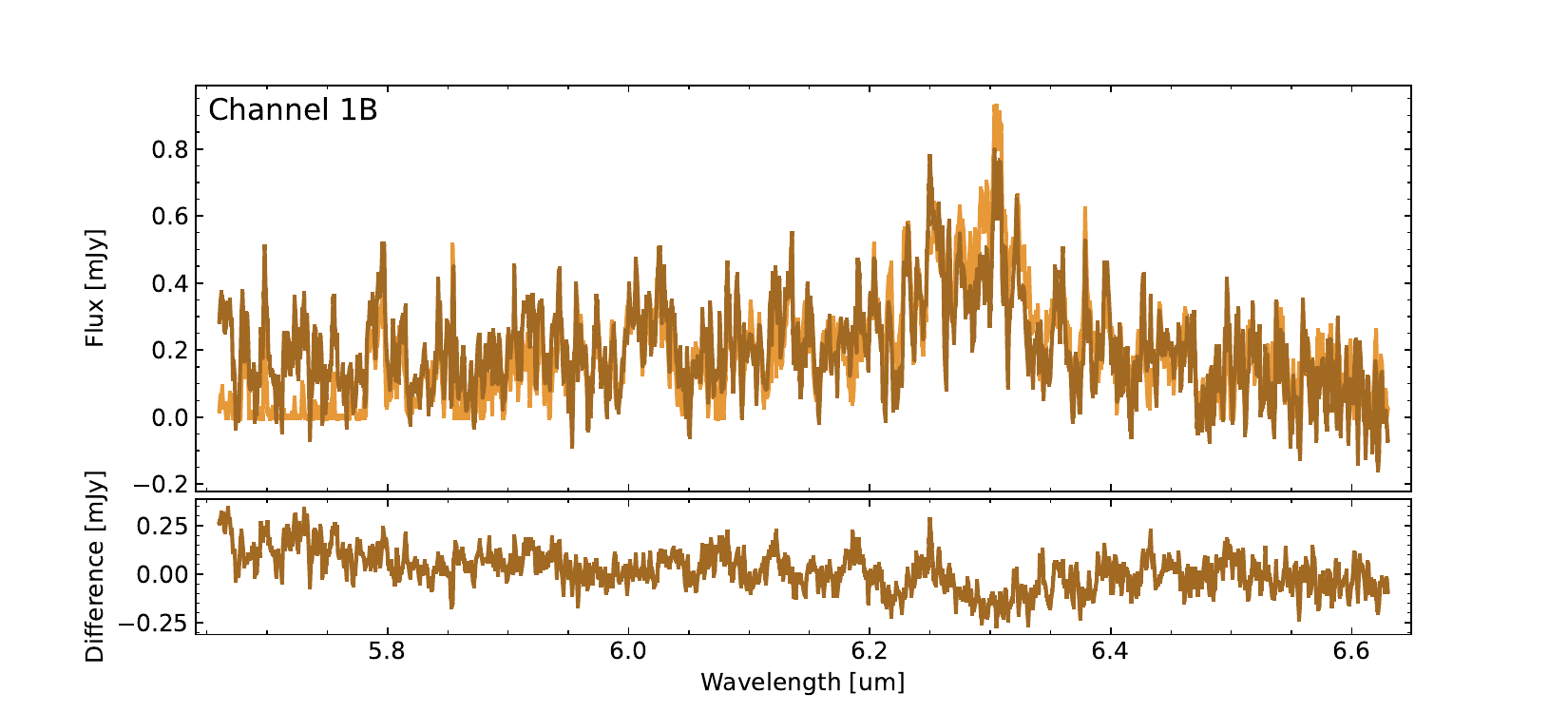}
    \includegraphics[width=0.8\textwidth, trim=1cm 0cm 2.3cm 1.1cm, clip]{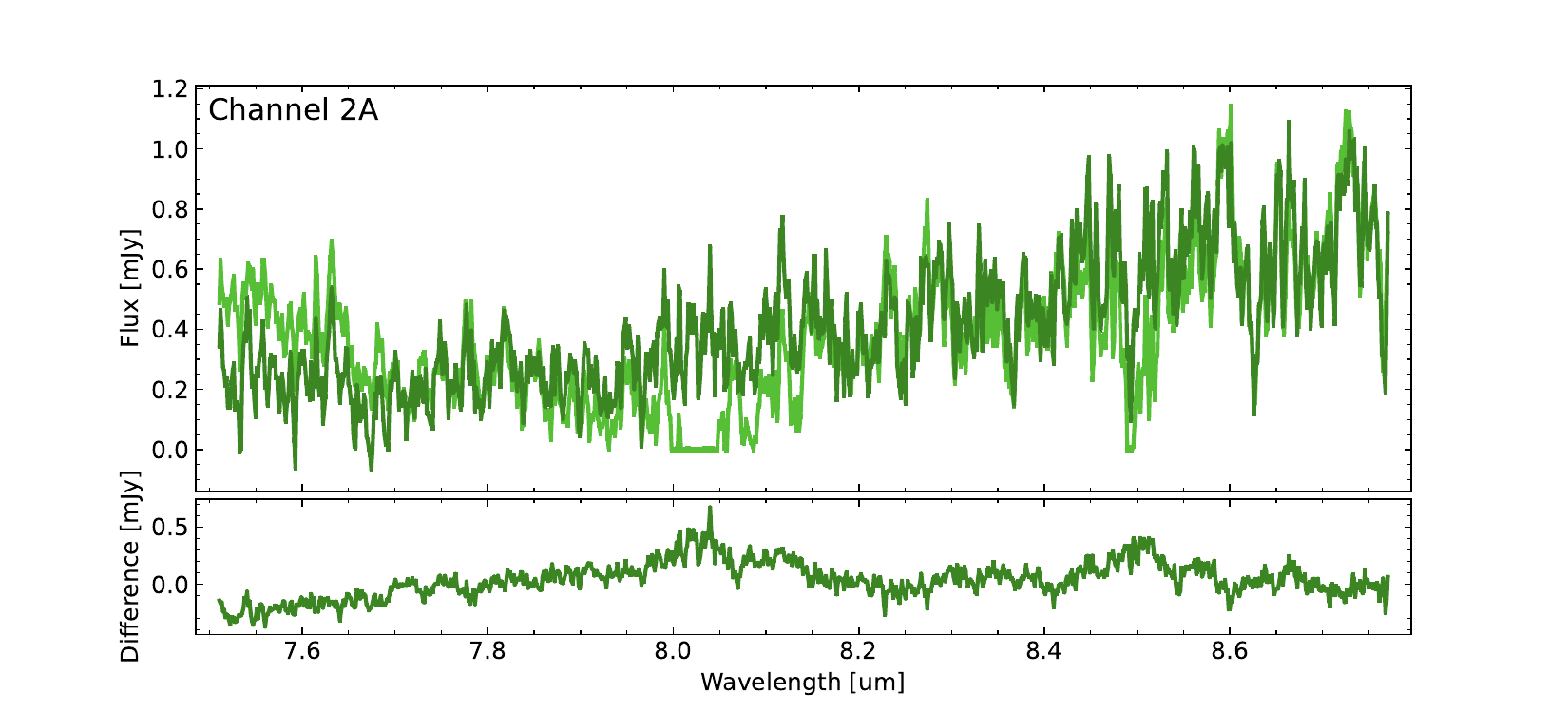}
    \caption{Spectrum of the target before (light colors) and after (dark colors) the residual background correction (Appendix \ref{app:backgrounds}). We show the full spectrum in the upper panel, and zooms of channel 1B and 2A in the middle and lower panels. The correction significantly impacts the overall shape of the spectrum, as well as altering the shapes of some features in the spectrum. In both cases the spectrum is extracted using an aperture radius of 2$\times$FWHM, to ensure that the flux of both components of the brown dwarf binary is captured. Each different color represents a different sub-channel of the MIRI MRS.}
    \label{fig:speccompare_backgrounds}
\end{figure*}

To correct for this effect, we performed a data-driven post-processing to the 3D cubes. The code for this correction is available at \url{https://github.com/ecmatthews/mirimrs_destripe}. We found that a relatively simple treatment was effective: in the \texttt{ifualign} orientation, we subtracted the mean value of each row and each column from that row/column, in order to ``flatten" the background and remove large-scale structures. We then used the standard \texttt{Extract1dStep} from the \texttt{jwst} pipeline to extract a spectrum from these ``flattened" \texttt{*s3d.fits} files. While this does not account for systematics that change within a row/column, we found that these are rarely present in the background of our frames. In Figure \ref{fig:speccompare_backgrounds} we highlight the impact of this correction on the extracted spectrum. The correction can impact the overall slope of the spectrum (e.g. around 7.6\,\micron), as well as introduce sharper features, and change the shapes of absorption features (e.g. around 8\,\micron). 

\begin{figure*}
    \centering
    \includegraphics[width=.94\textwidth]{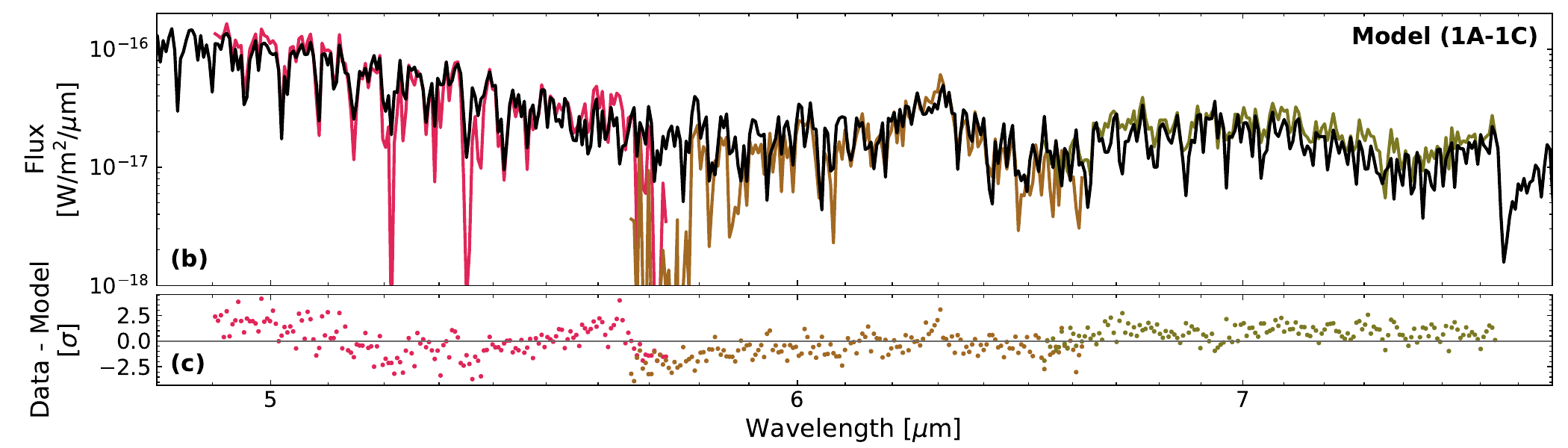}
    \includegraphics[width=.94\textwidth]{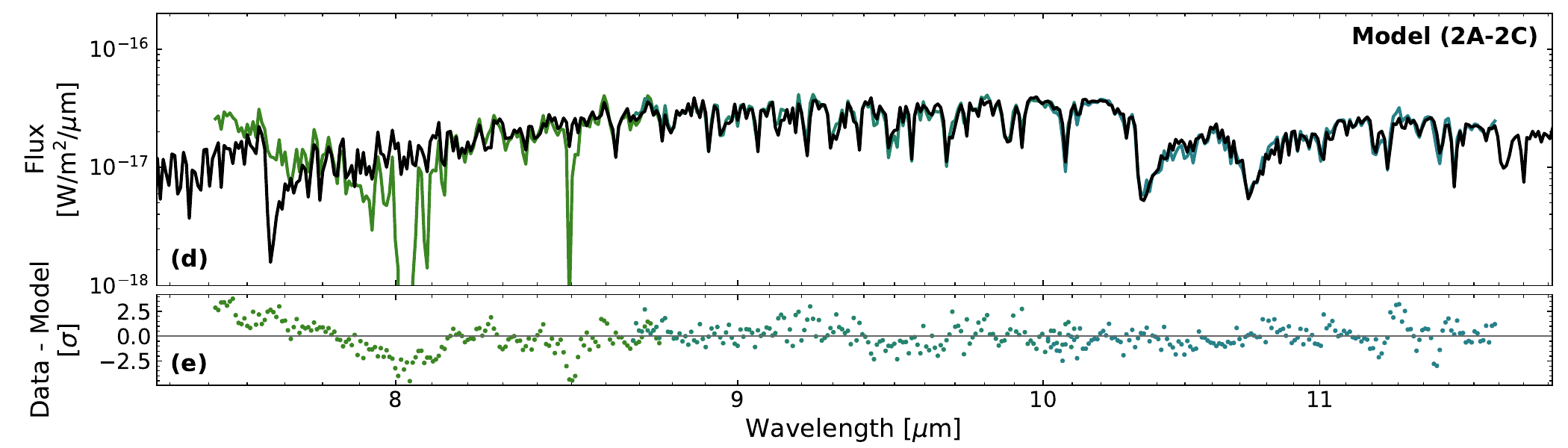}
    \includegraphics[width=.94\textwidth]{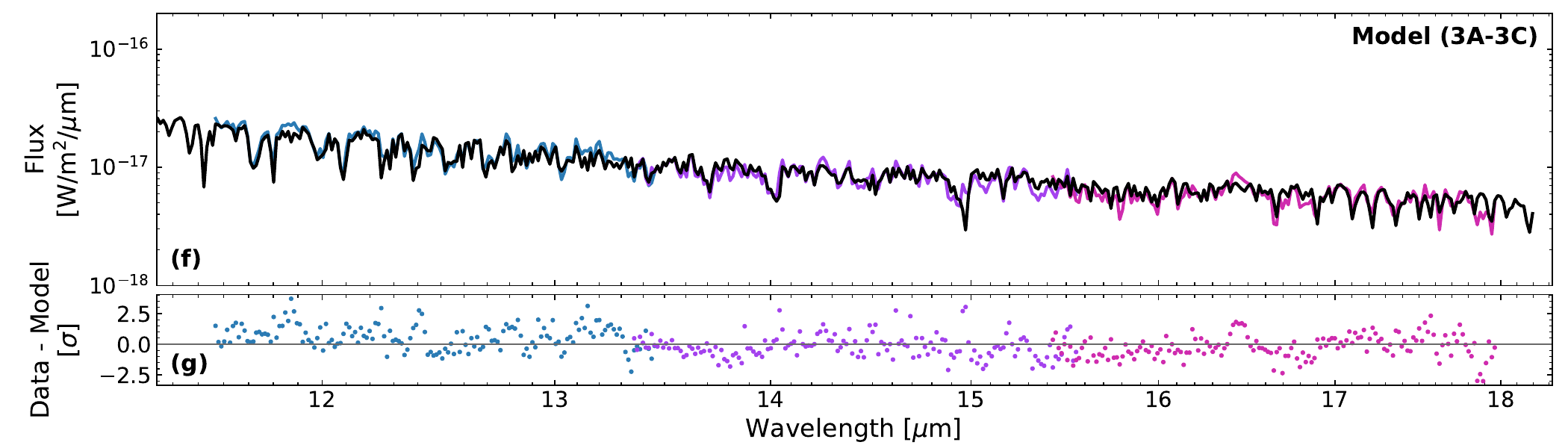}
    \caption{Same as Figure \ref{fig:mainspectrum_and_model} panels (b)-(g), but for the spectrum without the background correction applied (see Appendix \ref{app:backgrounds} for details). The dispersion of datapoints around the model is larger than in Figure \ref{fig:mainspectrum_and_model}, and wavelength-dependent systematics can clearly be seen in the residuals (lower panel) which show structured deviations from 0. Systematics impact the shape and presence of absorption features; retrieved abundances are correspondingly biased.}
    \label{fig:badspecfit}
\end{figure*}

These changes in turn bias the outcome of retrieval analyses on the measured spectrum. We illustrate this in Figure \ref{fig:badspecfit} with a retrieval on data without this background correction, where extended wavelength-dependent systematics are visible, with the same physical model as for Fig.~\ref{fig:mainspectrum_and_model}. There are clear, wavelength-dependent systematics remaining in the residuals (unlike in Fig.~\ref{fig:mainspectrum_and_model} above) that cannot be explained with an atmosphere model. Further, the shapes of key molecular features are distorted, and the retrieved abundances of molecules in the atmosphere are correspondingly biased.

Future works should explore more sophisticated corrections to this residual background term. In particular, a data-driven post-processing in the calibrated detector images produced by the stage 2 pipeline (\texttt{*cal.fits}) might allow for large-scale structures in the residual background term to be modeled and removed. We further recommend that several different dither positions be used when collecting spectra (as opposed to the two dither positions used in the present work) to provide a larger dataset with which to model and remove the background emission.

\end{document}